\documentclass[10pt,a4paper,twocolumn,tightenlines,amsmath,amssymb,nofootinbib,superscriptaddress,showpacs]{revtex4-1}

\usepackage{natbib}
\usepackage[hyperindex,breaklinks]{hyperref}
\usepackage{breakurl}
\usepackage{cleveref}

\usepackage{graphicx}
\usepackage[font={small,it}]{caption}
\usepackage{subcaption}
\usepackage{epstopdf}
\newcommand{\caphead}[1]{{\bf #1}}
\usepackage{stmaryrd}

\usepackage{bm}
\usepackage{bbm}

\usepackage{amsmath}
\usepackage{amsthm}
\usepackage{amssymb}

\usepackage{tabularx}
\usepackage{gensymb}

\newcounter{tlc}

\newtheorem{theorem}[tlc]{Theorem}
\newtheorem{lemma}[tlc]{Lemma}
\newtheorem{corollary}[tlc]{Corollary}
\crefname{tlc}{}{}

\newcounter{resultnum}

\newtheorem{result}[resultnum]{Result}
\crefname{resultnum}{Result}{Results}
\Crefname{resultnum}{Result}{Results}

\newtheorem*{example}{Example}

\usepackage{changepage}
\newcommand{\mainresult}[2]{\begin{adjustwidth}{1.5em}{1em} \emph{\cref{#1})} #2 \end{adjustwidth}}

\usepackage{enumitem}

\makeatletter
\newcommand*{\balancecolsandclearpage}{%
  \close@column@grid
  \clearpage
  \twocolumngrid
}
\makeatother

\usepackage{enumitem}
\allowdisplaybreaks[1]
\usepackage[super]{nth}

\makeatletter
\DeclareRobustCommand{\cev}[1]{%
  \mathpalette\do@cev{#1}%
}
\newcommand{\do@cev}[2]{%
  \fix@cev{+}%
  \reflectbox{$\m@th#1\vec{\reflectbox{$\fix@cev{-}\m@th#1#2\fix@cev{+}$}}$}%
  \fix@cev{-}%
}
\newcommand{\fix@cev}[1]{%
  \mathchoice%
    {\mkern#13mu}
    {\mkern#13mu}
    {\mkern#12mu}
    {\mkern#12mu}
}
\makeatother

\usepackage{stackengine}
\def\vecsign{\mathchar"017E}
\def\dvecsign{\smash{\stackon[-1.95pt]{\vecsign}{\rotatebox{180}{$\vecsign$}}}}
\def\dvec#1{\def\useanchorwidth{T}\stackon[-4.2pt]{#1}{\,\dvecsign}}
\stackMath

\newcommand{\past}[1]{\cev{#1}}
\newcommand{\future}[1]{\vec{#1}}
\newcommand{\pastfuture}[1]{\dvec{#1}}

\newcommand{\kB}{k_\mathrm{B}}

\providecommand{\Pr}{}
\renewcommand{\Pr}[1]{\mathrm{P}{\small(} { #1 } {\small)}}
\newcommand{\cPr}[2]{\mathrm{P}{\small(}#1 \,|\, #2{\small)}}
\newcommand{\Info}[2]{{I}{\small(}#1 \,; #2{\small)}}
\newcommand{\Ent}[1]{{H}{\small(}#1\small)}
\newcommand{\cEnt}[2]{{H}{\small(}#1 \,|\, #2{\small)}}

\newcommand{\inlineheading}[1]{\textbf{{#1}}}
\newcommand{\inlinesubheading}[1]{\textit{#1}}

\newcommand{\CQT}{Centre~for~Quantum~Technologies, National~University~of~Singapore, 3 Science Drive 2, 117543, Singapore}
\newcommand{\Tsing}{Center~for~Quantum~Information, Institute~for~Interdisciplinary~Information~Sciences,\\ Tsinghua~University, Beijing, 100084, China}
\newcommand{\Oxf}{Atomic~and~Laser~Physics, University~of~Oxford, Clarendon~Laboratory,\\Parks Road, Oxford, OX1 3PU, United Kingdom.}
\newcommand{\NUSPhys}{Department of Physics, National University of Singapore, 3 Science Drive 2, Singapore 117543}
\newcommand{\NTUPhys}{School of Physical and Mathematical Sciences, Nanyang Technological University, 639673, Singapore}
\newcommand{\NTUComplex}{Complexity Institute, Nanyang Technological University, 639673, Singapore.}

\begin{document}

\title{Thermodynamics of complexity and pattern manipulation}
\author{Andrew~J.~P.~Garner}
\email{ajpgarner@nus.edu.sg}
\affiliation{\CQT}
\affiliation{\Tsing}

\author{Jayne Thompson}
\affiliation{\CQT}

\author{Vlatko Vedral}
\affiliation{\Oxf}
\affiliation{\CQT}
\affiliation{\NUSPhys}
\affiliation{\Tsing}

\author{Mile Gu}
\email{cqtmileg@nus.edu.sg}
\affiliation{\NTUPhys}
\affiliation{\NTUComplex}
\affiliation{\CQT}
\affiliation{\Tsing}

\date{\today}

\begin{abstract}
Many organisms capitalize on their ability to predict the environment to maximize available free energy, and reinvest this energy to create new complex structures.
This functionality relies on the {\em manipulation of patterns}---temporally ordered sequences of data.
Here, we propose a framework to describe pattern manipulators -- devices that convert thermodynamic work to patterns or vice versa --  and use them to build a `pattern engine’ that facilitates a thermodynamic cycle of pattern creation and consumption.
We show that the least heat dissipation is achieved by the provably simplest devices; the ones that exhibit desired operational behaviour while maintaining the least internal memory.
We derive the ultimate limits of this heat dissipation, and show that it is generally non-zero and connected with the pattern’s intrinsic {\em crypticity} -- a complexity theoretic quantity that captures the puzzling difference between the amount of information the pattern's past behaviour reveals about its future, and the amount one needs to communicate about this past to optimally predict the future.
\end{abstract}
\maketitle

\begin{figure}[tbh]
\includegraphics[width=0.425\textwidth]{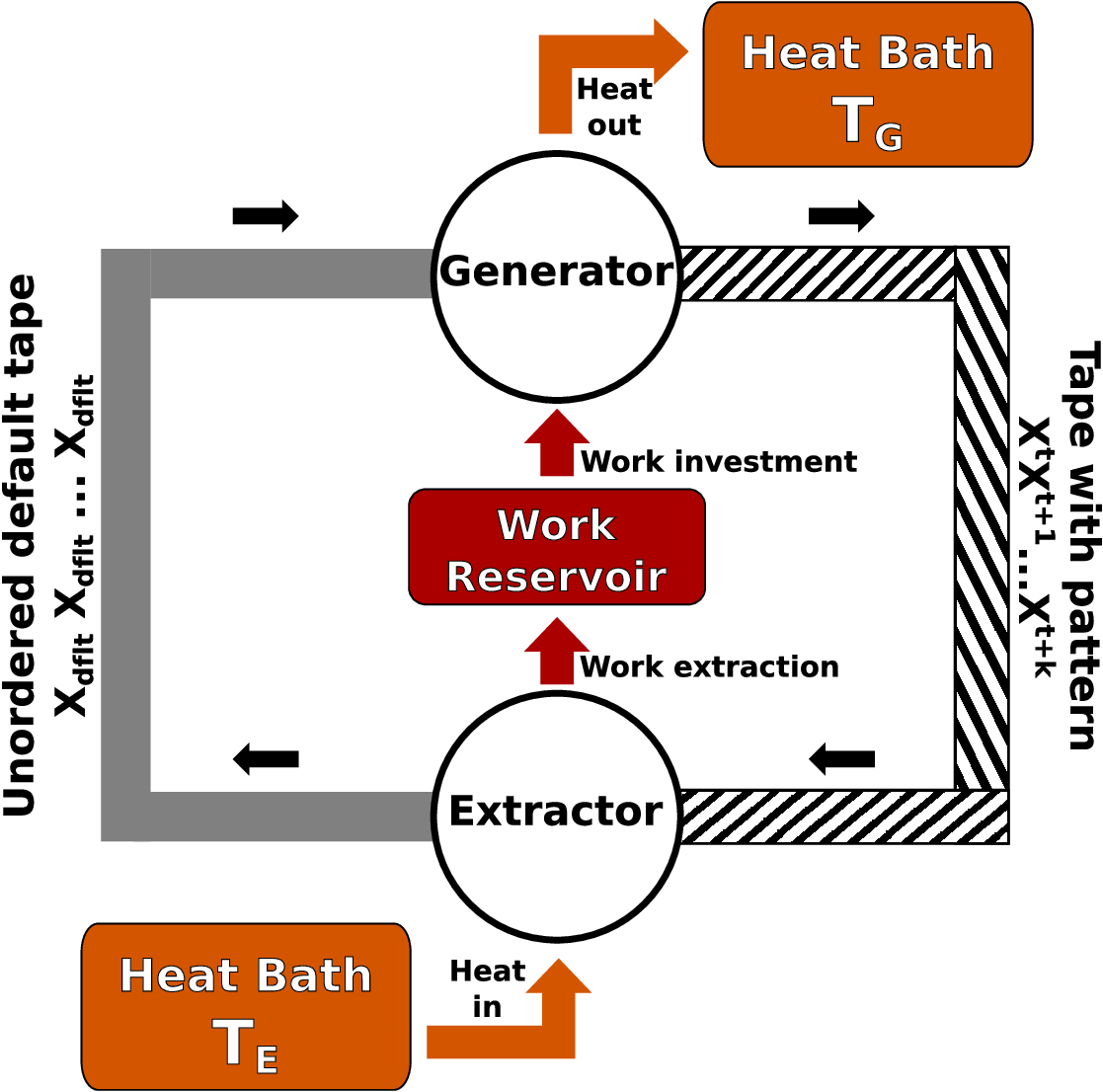}
\caption{
\label{fig:SimExtCycle}
\caphead{Cycle of pattern generation and extraction.}
A tape moves through the system in a clockwise manner.
A generator expends work to write a pattern to the tape.
The extractor then uses the pattern on this tape to extract work.
To run cyclically, each device maintains {\em prescient} memory that keeps track of the pattern.
In this article, we identify the dissipative work costs.
We find that the simplest generator has the best thermodynamic performance;
 but surprisingly for the extractor, the choice of memory has no thermodynamic consequence.
}
\end{figure}

The manipulation of patterns is as important to living organisms as it is for computation.
Living things capitalize on structure in their environment for available energy, and use this energy to generate new complex structures.
Similarly, a crucial task in the modern era of big data is to identify patterns in large data sets in order to make predictions about future events -- often at great energetic cost. Here, we consider the thermodynamic costs intrinsic to this sort of pattern manipulation and ask: is there a preferred method by which this manipulation should be done?
Our intuition is that {\em simpler is better}, a longstanding tenant of natural philosophy known as Occam's razor. To formalize this, we first qualify what is meant both by \emph{simpler} and by \emph{better}.

In complexity science, computational mechanics formalizes what is {\em simpler} in the context of pattern manipulation~\cite{CrutchfieldY89,ShaliziC01,Crutchfield11}. 
The premise is that everything we observe in the environment can be considered to be a pattern --- a temporal sequence of data exhibiting certain statistical structure. 
Much of science then deals with building models that can explain such statistics -- machines that take information from past observations, and use it to generate statistically coinciding conditional future predictions. 
Given two machines that exhibit same pattern of behaviour, the one that stores less information from the past is considered simpler, the motivation being that it better isolates indicators of future behaviour. 
The simplest such machine then defines exactly how much memory is required to produce a given pattern, and thus quantifies the pattern's intrinsic structure. 
Known as {\em statistical complexity}, this measure has been applied to quantify structure in diverse contexts~\cite{HaslingerKS10,ParkWLYJM07,LuB12}.

Meanwhile in thermodynamics, \emph{better} originally described heat engines that produce more work with less wasted heat. 
This carries through to modern thermodynamics:  the best approach for a given task being the one that minimizes the expenditure of a limited resource~\cite{BrandaoHORS13,HorodeckiO13} (e.g.\ work or hard-to-create states).
 Since information is physical~\cite{Bennett82,Landauer96,LeffR02,ParrondoHS15}, patterns, which are correlated information, are also physical and hence subject to the laws of thermodynamics. 
 In this context, a pattern may be treated as an {\em information reservoir} – a source of free energy encoded in correlations~(e.g.\ \cite{Bennett82,MandalJ12,DeffnerJ13,BaratoS14,ParrondoHS15,Wolpert15,BoydMC16,BoydMC16a,BoydMC16b}). 
{\em Pattern manipulators}, which convert a pattern to useful work or vice versa, are thus a type of heat engine. 
The pattern manipulator that effects a prescribed change in a pattern with the minimal heat dissipation can thus be regarded as {\em better}.

Here, we derive the fundamental thermodynamic limits for the manipulation of patterns by devices operating in a cycle (see~\cref{fig:SimExtCycle}).
We do not re-derive the second law, but (believing it highly likely to be true) consider the implications that it places on the work dissipation intrinsic to {\em all} possible patterns manipulators. 
Our approach is to connect pattern manipulators in the context of thermodynamics to predictive models in the context of computational mechanics -- observing that the creation or consumption of a given pattern involves retaining enough of the past to correctly anticipate its expected future statistics.
We show that simplest patterns manipulators  (i.e.\ ones that store the least information about the past of the pattern) results in the least dissipation--- and thus simpler is thermodynamically better.

We describe the simplest, most thermodynamically efficient causal pattern manipulators -- those whose memory requirement is given by the pattern’s statistical complexity -- and show that they, remarkably, still must dissipate some excess heat.
We show that this heat dissipation is lower-bounded by the {\em crypticity} of the pattern~\cite{CrutchfieldEM09}, a hitherto complexity--theoretic property quantifying the puzzling difference between the amount of information the past of a pattern reveals about its future, and the amount of information one needs to communicate about the past of the pattern in order to predict its future.
These bounds apply to any model in any physical framework that can implement pattern manipulation tasks described in a manner consistent with Landauer's principle and computational mechanics.
Our work thus highlights the many thermodynamic consequences of complexity in pattern manipulation.

\inlineheading{Patterns as a resource.}
Knowing a system's internal state has thermodynamic consequence.
This knowledge can be used to perform work (drive a mechanical task),
 as illustrated by the Szil\'ard engine thought experiment~\cite{LeffR02}:
A box has a single particle inside, on the left- or right-hand-side.
A movable barrier inserted in the box's center acts as a piston that expands as the particle pushes against it.
If an agent knows which side of the barrier the particle is on, she can couple the barrier (e.g.\ via a pulley) to raise a weight.
As the piston expands, it lifts the weight and generates an amount of work $\kB T \ln 2$ at temperature $T$ by drawing in the same amount of heat from its surroundings.

Knowledge about patterns may also be exploited. Consider a sequence of Szil\'ard engines arranged in a linear configuration on a conveyer belt (or more abstractly, symbols on a tape), indexed sequentially by $t\in\mathbb{Z}$.
These engines are prepared such that the particle in engine $t$ is on the same side of the barrier as the in engine $t\!-\!1$ with probability $p\neq\frac{1}{2}$.
Suppose now that an agent attempts to extract work from these engines in sequential order.
An agent unaware of this pattern would only be able to correctly predict the particle's location half of the time,
 and hence will extract less work than an agent who knows the pattern and couples her pulley accordingly.
As such, the ability to predict grants thermodynamic advantage.

This is a manifestation of Maxwell's d\ae{}mon---an apparently paradoxical conversion of heat into work 
 that is only resolved by accepting that information is physical and hence subject to the laws of thermodynamics~\cite{Bennett82,Landauer96,LeffR02,ParrondoHS15}.
For the single Szil\'ard engine, we must also account for the cost of resetting the agent's memory about the particle's location---this knowledge must be thought of as a resource.
Likewise, since it is more thermodynamically useful that the sequence follows a pattern than be uncorrelated,
 the pattern itself must also be considered as a resource.
Producing a pattern hence requires an investment of work.
Moreover, any physical device that generates (or exploits) a pattern contains some memory about what has been observed in the pattern so far,
 in order to accurately generate (or anticipate) upcoming parts of the pattern.
Any thermodynamic costs of maintaining this internal memory must also be accounted for.
One thus can think of a tape with a pattern as one thinks of a rising weight (or any other form of battery) and treat it as an {\em information reservoir}~\cite{Bennett82,MandalJ12,DeffnerJ13,BaratoS14,ParrondoHS15,Wolpert15}.
This approach has been used recently as a bridge from computational mechanics to modern thermodynamics~\cite{BoydMC16,BoydMC16a,BoydMC16b}.

The quantitative link between information, entropy and heat dissipation is given by Landauer's principle~\cite{LeffR02}:
 the minimum work cost of any information-processing task is proportional to the total change in information entropy\footnote{
 Classically, given by the {\em Shannon entropy} $\Ent{X} = -\sum_i \Pr{X=x_i} \log_2 \Pr{X=x_i}$.
Using a base 2 logarithm gives units of {\em bits}.
}
 (just like macroscopic thermodynamics, where the minimum work required to slowly change between two states of the same internal energy is proportional to the change in thermodynamic entropy).
%

Landauer's principle sets a lower bound on the work cost of performing an information processing task.
In this article, we shall take Landauer's principle as our starting point. 
Our results have meaning in any physical framework that: 1.\ provides a way of mapping random variables to physical states, and 2.\ has definitions of heat and work exchange, such that any allowed reconfiguration of these mapped states has a work cost bounded by Landauer's principle, as applied to the associated random variables.
A list of such frameworks satisfying these criteria includes the trajectory formalism~\cite{Alicki79,AlickiHHH04,Kieu06,QuanD08,KimSDU11,BrowneGDV14} (see {\em Physical Example} in the {\em Technical Appendix}), the resource theory of thermodynamics (e.g.\ \cite{BrandaoHORS13,HorodeckiO13}), single-shot statistical mechanics (e.g.\ \cite{Aberg13,YungerHalpernGDV15,EgloffDRV15}), among others.

In these frameworks, Landauer's principle is not an additional imposition on top of the physics, but rather manifests from microdynamical behaviour -- it is an emergent law, much like the second law of thermodynamics.
It is not necessary to know the particulars of the microdynamics to derive or apply the results in this article --
 so long as the framework obeys Landauer's principle, any model in that framework that implements the pattern manipulations presented in the following sections will be bound by our results.

\inlineheading{The framework of patterns.} Computational mechanics provides a formal framework for describing patterns \cite{CrutchfieldY89,ShaliziC01,CrutchfieldEM09,Crutchfield11}. 
Consider a sequence of physical systems indexed by a parameter $t\in\mathbb{Z}$, each having a configuration space $\{x\}$. A general pattern on such a sequence of systems is defined by a bi-infinite sequence of random variables $\pastfuture{X} = \ldots X^{t-1} X^t X^{t+1} \ldots$, where $X^t$ governs the configuration of system $t$. Here, one generally considers devices that observe the systems in some sequential order, such that system $t$ is observed at time $t$. Thus, $\past{X}^t = \ldots X^{t-1}X^t$ govern the configurations of all systems that could have been observed in the past, and $\future{X}^t =  X^{t+1}X^{t+2}\ldots$ governs the configurations of all systems that could be observed in the future.
A statistical description of the {\em pattern} is given by the probability distribution $\Pr{\past{X}^t,\future{X}^t}$, where each instance of such a pattern taking on a particular configuration $\pastfuture{x} = \ldots x^{t-1} x^t x^{t+1} \ldots$ occurs with probability  $\Pr{\pastfuture{X} = \pastfuture{x}}$.

When there is a meaningful pattern, the past and future are correlated.
For a particular instance of the pattern where $\past{x}$ is observed, the statistics of future observations are given by $\cPr{\future{X}^t}{\past{X}^t=\past{x}}$.
The mutual information%
\footnote{
The {\em mutual information} between random variables $A$ and $B$ with domains $\mathcal{A}$ and $\mathcal{B}$ respectively, is given $\Info{A}{B} = \Ent{A} - \cEnt{A}{B}$ where $\cEnt{A}{B} := -\sum_{b\in\mathcal{B}}\Pr{B}\sum_{a\in\mathcal{A}} \cPr{A=a}{B=b}\log_2 \cPr{A=a}{B=b}$ is the {\em conditional entropy} of $A$ given $B$.
Using logarithms of base $2$, all have units of bits.
}
$\Info{\past{X}^t}{\future{X}^t} = \Ent{\future{X}^t}-\cEnt{\future{X}^t}{\past{X}^t}$ then quantifies how useful this past knowledge is for predicting the future. In computational mechanics, this value is known as the {\em excess entropy} $E$ (see definition 13 of~\cite{ShaliziC01}).

We make the simplifying assumption that the pattern is a {\em stationary} -- the statistics $\Pr{\past{X}^t,\future{X}^t}$ are invariant under time translation. This does not mean that every output $x^t$ in the sequence is identical, or that the pattern is Markovian,
but rather that the statistics of $X^{t+1}$ onwards, given a past sequence, have no explicit time dependence:
 $\cPr{\future{X}^t}{\past{X}^t=\past{x}} = \cPr{\future{X}^{t'}}{\past{X}^{t'}=\past{x}}$ for all $t$, $t'$ and $\past{x}$.
Hence, we can often omit the superscript~$t$.

\inlineheading{Cyclic pattern manipulation.}
The classical expressions of thermodynamic laws (e.g.\ Kelvin's statement that a device cannot convert heat into work with no other effect~\cite{Thomson51}) concern cyclic behaviour---%
 processes that leave the system in a state allowing for repetition with the same thermodynamic consequences.
Without a full cycle in mind, there is the danger that the thermaldynamic benefit of a process may come at the expense of consuming an unaccounted-for resource.
A cycle does not require the {\em microstate} of the system to return to its original value.
Consider a piston of gas expanding and compressing:
 it does not matter if the individual molecules have moved to new locations by the end of the cycle, as long as the important thermodynamic variables---the pressure and volume---return to their original values.

We now establish a framework for understanding the thermodynamics of generating and consuming patterns as a cyclic process (\cref{fig:SimExtCycle}).
Consider a sequence of physical systems indexed by a parameter $t\in\mathbb{Z}$, each having a configuration space $\{x\}$.
We assume that the default configuration for these systems to be uncorrelated, such that the configuration of each system is governed by some default random variable $X_{\rm dflt}$.
The mathematical results of this article will hold for any choice of $X_{\rm dflt}$, though for clarity, we shall frame our discussion as if $X_{\rm dflt}$ represents the uniformly random distribution.

For each run of the cycle, we act on a moving window of length $k$ on this sequence.
We index the first system in this window as $t+1$, and the final system as $t + k$.
Between each run of the cycle, the window advances by $k$ systems, such that the subsequent cycle acts on systems $t + k + 1$ to $t + 2k$.
We refer to $k$ as the {\em stride} of the cycle.

In each cycle, the tape is acted on by two machines (\cref{fig:SimExtCycle}) associated with the same pattern $\Pr{\past{X},\future{X}}$.
Each machine is in contact with a thermal reservoir (i.e.\ heat bath at inverse temperature $\beta=\frac{1}{\kB T}$) and a battery for storing free energy (e.g.\ a raising weight).
The first machine, a {\em generator}, does work in order to act on the systems in the window $t+1, t+2, \ldots, t+k$, taking their configurations from ${X_{\rm dflt}}^{\otimes k}$ to configurations governed by the random variables $X^{t+1}X^{t+2}\ldots X^{t+k}$ from the pattern.
The second machine, an {\em extractor}, resets these $k$ systems back to their uncorrelated default state ${X_{\rm dflt}}^{\otimes k}$, outputting work as it does so.
Collectively, we refer to these devices as {\em pattern manipulators}.

Each of these two devices can operate independently.
By itself, the generator operating ad infinitum encodes an arbitrarily long section of the pattern onto a sequence of physical systems that were initially in the default configuration. This will require an investment of work, and produces a pattern as a resource.
Likewise, the extractor consumes a section of the pattern of arbitrary length, extracting work and resetting the systems back to their default configuration.
However, to account for all thermodynamic resources, it is helpful to consider them operating together as the cycle described above (or in \cref{fig:SimExtCycle}).

\inlineheading{Prescient memory.}
For any device to generate a particular pattern $\pastfuture{X}$ in time-sequential order, it must contain an internal memory (with some configuration space $\{r\}$) that records some information about what it has generated before. 
Consider any pattern with two possible pasts $\past{x}$ and $\past{x}'$ that yield differing conditional future statistics; i.e., $\cPr{\future{X}}{\past{X}\!=\!\past{x}} \neq \cPr{\future{X}}{\past{X}\!=\!\past{x}'}$.
A machine that generates such a pattern must behave differently depending on whether it has generated $\past{x}$ or $\past{x}'$ so far. 
Thus the state of the machine's memory $\{r\}$ must be dependent on $\past{x}$. 
Likewise, an extractor must also adjust its future actions based on what past $\past{x}$ it has observed so far to best harness the free energy in $\future{x}$. 
Thus, it must also field some $\past{x}$-dependent memory.

Each particular strategy for recording this past information can be captured by some mapping $f$ that describes the state of the memory, $r$, depends on $\past{x}$. $f$ is referred to as an \emph{encoding map}, and defines a particular strategy in which a given generator or extractor keeps track of the past.
The encoding map then induces a probability distribution over $r$, governed by some random variable~$R$.

To generate a desired pattern governed by $\Pr{\past{X},\future{X}}$, the memory $R$ must be {\em prescient} with respect to the pattern.
That is, for all possible $\past{x}$,
 $R$  satisfies  $\cPr{\future{X}}{\past{X}\!=\!\past{x}} = \cPr{\future{X}}{R\!=\!r}$,  for any $r$ assigned by $f\!\left(\past{x}\right)$. 
In words: for the purpose of inferring or generating the pattern's future statistics behaviour $\future{X}$, knowing the state of {\em prescient memory} $r$ is as useful as knowing the entire semi-infinite string of past outcomes $\past{x}$.
It follows that for any prescient memory $R$, the mutual information $\Info{R}{\future{X}} = \Info{\past{X}}{\future{X}}$.
We shall consider only memory that acts in a {\em causal way}, such that the state of the memory at time $t$ does not contain any information about the future of the pattern $\future{X}$, that is not already contained in the past. (In the language of computational mechanics, we do not allow the memory to have {\em oracular information}~\cite{CrutchfieldEJM10}).

Given any pattern, there are clearly many encoding maps that generate memory states satisfying prescience.
The most obvious choice is the identity map, which represents a generator that remembers every single step of the pattern it has generated up to the present.
Such choices are clearly wasteful.
Formally, the Shannon entropy of the memory, $\Ent{R}$ would coincide with $\Ent{\past{X}}$, and be unboundedly large -- even when generating patterns with no structure (e.g. a completely random sequence).
This has motivated complexity theorists to look for more efficient encodings -- corresponding to generators whose corresponding memory, $H(R)$, is minimized.

The {\em causal states}~\cite{CrutchfieldY89,ShaliziC01} represent the most efficient prescient encoding.
They coincide with the equivalence classes defined by the equivalence relation $\sim_\epsilon$, where $\past{x} \sim_\epsilon \past{x}'$
 if and only if $\cPr{\future{X}}{\past{X} = \past{x}} = \cPr{\future{X}}{\past{X} = \past{x}'}$. 
The optimal machine does not store which past $\past{x}$ has occurred, but rather which equivalence class $s$ that $\past{x}$ belongs to -- the rationale being that two pasts in the same equivalence class do not need to be distinguished as they have coinciding conditional futures. 
This motivates a particular encoding map $\epsilon$ that deterministically takes each $\past{x}$ to a particular causal state $s$ whenever $x \in s$. 
The resulting random variable $S$ over causal states is uniquely defined for each pattern $\pastfuture{X}$, and represents the prescient memory with minimal entropy~\cite{CrutchfieldY89,Shalizi01}. 
The memory required to store these states, given by the Shannon entropy $C_\mu = H(S)$, is known as the pattern's \emph{\mbox{statistical} complexity} and quantifies the pattern's intrinsic structure.

\begin{figure}[hbt]
\includegraphics[width=0.375\textwidth]{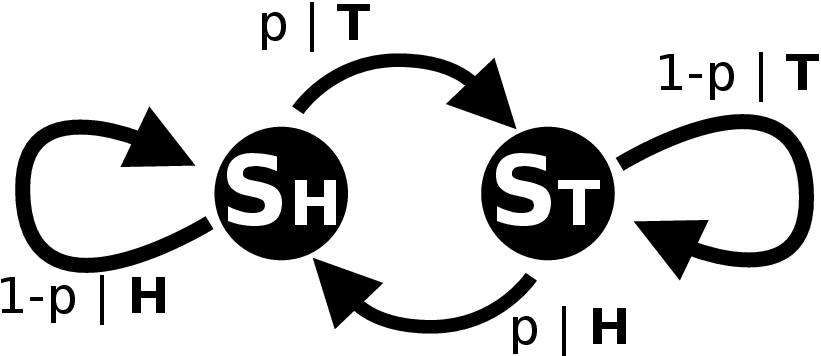}
\caption{
\label{fig:PerturbedCoin}
\caphead{Perturbed coin pattern  $\epsilon-$machine ($k=1$).}
The ``perturbed coin'' pattern may be produced by a weighted random walk on this network.
The nodes represent the two configurations of the $\epsilon-$machines internal memory $R\in\{s_h, s_t\}$ (i.e.\ the {\em causal states}).
The directed edges represent the effect of a single time step: a (possible) change in memory configuration, and the setting of the system on the tape to some value.
The labels $P \,|\, \mathbf{x}$ gives the probability $P$ of transitioning to the particular memory configuration $R^{t+1}$ (conditional on the initial memory configuration $R^t$) while configuring the system at index $t+1$ on the tape into the state ${\bf x} \in \{H, T\}$).
}
\end{figure}

 In computational mechanics, generators and predictors that store only the causal states are considered the simplest such devices, and known as $\epsilon$-machines. Furthermore any pattern can be described by dynamics on causal states (see \cite{JamesMEC14} or our subsequent example in \cref{fig:PerturbedCoin}), and significant work exists on inferring causal states and the resulting $\epsilon$-machines from raw observational data~\cite{CrutchfieldY89,Shalizi01,ParkWLYJM07,HaslingerKS10}. 
Throughout this article, we use $S$ to represent the random variable governing the pattern's causal state, and $R$ to represent the random variable governing the state of some generic choice of prescient memory.
$R$ could be in one-to-one correspondence with $S$ (since $\epsilon$ is a specific choice of encoding map $f$), but in general it does not have to be.

\inlineheading{Prescient pattern manipulators.}
To operate continually in a cycle,
 a pattern manipulator's memory must be prescient before and after the device has acted on the pattern.
Suppose a generator at time $t$ starts with memory in configuration $R^t=r_i$.
After encoding $k$ systems into the pattern $X^{t+1}\ldots X^{t+k} = x^{t+1}\ldots x^{t+k}$,
 the final state of memory $R^{t+k} = r_j$ must satisfy
 $\cPr{\future{X}^{t+k}}{{R^{t+k}\!=\!r_j}} = \cPr{\future{X}^{t+k}}{R^t\!=\!r_i, X^{t+1}\ldots X^{t+k}\!=\!x^{t+1}\ldots x^{t+k}}$.
We call this condition {\em maintaining prescience}, since it implies that $R^{t+k}$ must also be prescient for the entire history of the pattern to that point, including the variate $x^{t+1}\ldots x^{t+k}$ just produced.
Likewise, if an extractor encounters and resets $X^{t+1}\ldots X^{t+k} = x^{t+1}\ldots x^{t+k}$ on the tape,
 it must also update its memory from $R^t$ to $R^{t+k}$ in a way that satisfies the same condition.

For a generator and extractor that maintain prescient memory $R$ and $R'$ respectively,
 if the initial configurations $r_i$ and $r_i'$ satisfy
 $\cPr{\future{X}^t}{R^{t}=r_i} = \cPr{\future{X}^t}{{R'}^{t}=r_i'}$,
 then the final configurations $r_j$ and $r_j'$
 (after $k$ steps of the pattern have been written and subsequently reset)
 will satisfy
 $\cPr{\future{X}^{t+k}}{R^{t+k}=r_j} = \cPr{\future{X}^{t+k}}{{R'}^{t+k}=r_j'}$.
This follows almost immediately from the definition of maintaining prescience (see lemma~\cref{lem:InStep} in the {\em Technical Appendix} for a proof).
In words:
 at the beginning of each run of the cycle,
 the portion of the pattern that the generator anticipates to next produce
 remains in alignment with the portion of the pattern that the extractor anticipates to next consume.

We have thus specified the action of pattern manipulators in terms of the initial and final states of their related information variables (i.e.\ state of the tape and state of the internal prescient memory).
In the following sections, we provide bounds on the minimum work costs of {\em any} process in any framework consistent with Landauer's principle that implements manipulations according to this specification.

\inlineheading{Investing work to generate a pattern.}
For any given pattern, there is a family of generators,
 characterized by their choice of prescient memory $R$, and by the number of steps $k$ of the pattern that they generate at once. 
In theorem~\cref{thm:generate} of the {\em Technical Appendix}, we prove that for any such generator, the work investment $W^k_{\rm gen}(R)$ required to generate $k$ steps of the pattern when the tape is subject to a degenerate Hamiltonian (that is, all configurations of the tape have equal energy), and the tape-memory system is coupled to a heat bath with inverse temperature $\beta=\frac{1}{\kB T}$, is bounded by:
\begin{align}
\label{eq:GenerateCost}
\beta W^k_{\rm gen}(R) \geq~ & k \left[ \Ent{X_{\rm dflt}} - \cEnt{X^{t+1}}{S^{t}} \right] \nonumber \\
& + (\cEnt{R^{t}}{X^{t+1}\ldots X^{t+k} R^{t+k}} \nonumber \\
& \quad - \cEnt{R^{t+k}}{R^t X^{t+1}\ldots X^{t+k}}),
\end{align}
where $S^t$ are the {\em causal states} of the pattern.
The bound holds regardless of what physical mechanism is used to generate the pattern, and relies only on the assumption that the generator obeys  Landauer's principle.
Equality is achieved when the theoretically optimal mechanism is used.
If the tape's Hamiltonian is not degenerate, the minium work investment requires an additional contribution $\Delta E := k \left[E(X) - E(X_{\rm dflt})\right]$, where $E(X)$ is the expectation value of a tape system's energy when configured into the pattern $X^t$, and $E(X_{\rm dflt})$ is the expectation value of the tape system's energy in its default state $X_{\rm dflt}$.

\begin{figure}[hbt]
\includegraphics[width=0.46\textwidth]{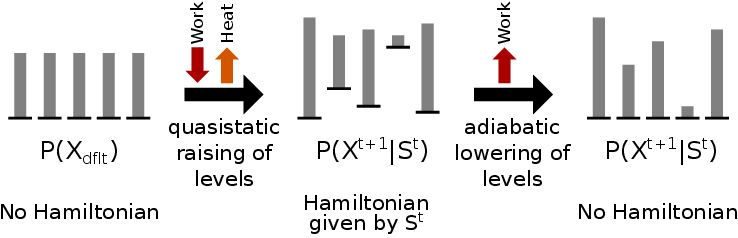}
\caption{
\label{fig:UpdateTape}
\caphead{One method of writing a pattern to a tape.}
The various choices of symbols on a tape can each be associated with a different energy level of a system (drawn as a black horizontal lines whose relative height indicates relative energy).
The statistical state of the symbol is a probability distribution (drawn as grey bars) over these configurations.
By changing the Hamiltonian of the tape whilst remaining in contact with a thermal reservoir, the statistics can be altered to $\cPr{X^{t+1}}{S^t}$.
At this point, the system is isolated from the heat reservoir and the Hamiltonian is adiabatically removed.
The whole procedure requires an investment of work proportional to the reduction in the state's entropy.
(See also {\em physical example} in the {\em Technical Appendix}.)
}
\end{figure}

We can divide the work cost of generating a pattern into two contributions: one from changing the entropy \emph{(and thus the free energy)} of the systems on the tape, and one from updating the internal memory.
The cost $W_{\rm tape}^k$ associated with writing $k$ symbols onto the tape (at inverse temperature $\beta = \frac{1}{\kB T}$) is given by the first line of \cref{eq:GenerateCost}:
\begin{equation}
\label{eq:TapeCost}
\beta \, W_{\rm tape}^k \geq k \left[ \Ent{X_{\rm dflt}} - \cEnt{X^{t+1}}{S^{t}} \right].
\end{equation}
This value is determined by the distribution over causal states~$S$ (recall that this is unique for any given pattern~\cite{ShaliziC01}) rather than the  specific device-dependent internal memory $R$.
As such, $W_{\rm tape}^k$ has no dependence on the choice of prescient memory, and scales trivially with the stride $k$, and is therefore an intrinsic property of the pattern rather than of the machine generating it (see \cref{fig:UpdateTape} for one possible physical realization of this cost using the framework of e.g.\ \cite{Aberg13}, or the ``physical example'' in the {\em Technical Appendix}).
(If the systems onto which the pattern is encoded do not have a degenerate Hamiltonian, the additional energy term $\Delta E$ should also be associated with the tape, and incorporated into $W_{\rm tape}^k$).

\begin{figure}[bth]
\includegraphics[width=0.45\textwidth]{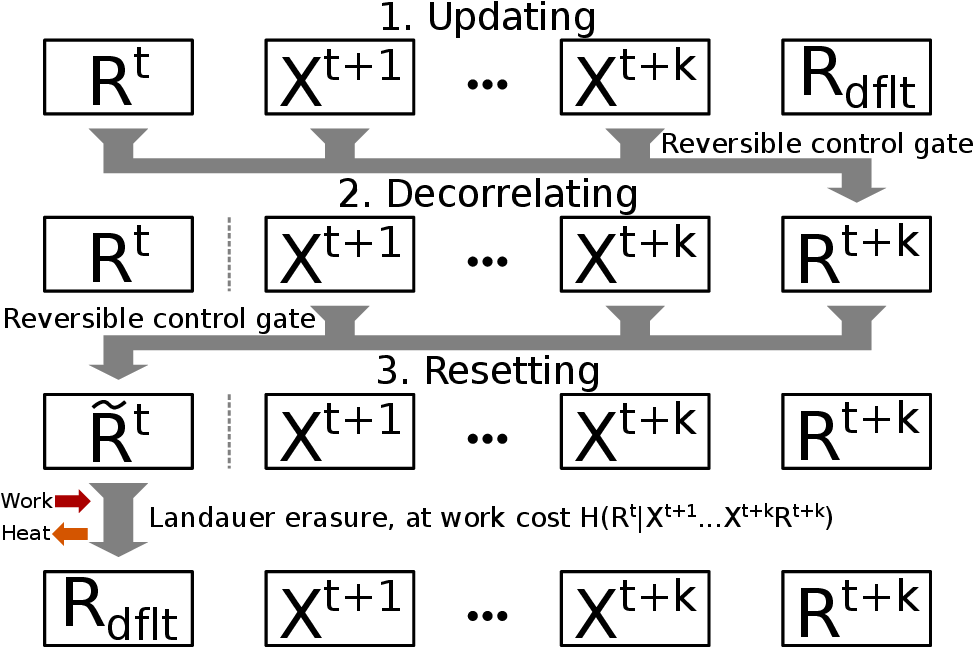}
\caption{
\label{fig:UpdateState}
\caphead{One method of updating the generator's memory.}
(The case for unifilar $R$ is shown.)
A blank pure ancilla state $R_{\rm dflt}$ is {\em updated} at no cost to $R^{t+k}$, conditioned on the values of the initial internal state $R^{t}$ and systems on the tape $X^{t+1}\ldots X^{t+k}$.
The old state $R^{t}$ is then {\em decorrelated} from $R^{t+1}$ and $X^{t+1}\ldots X^{t+k}$, and the mutual information used to reduce the entropy of the internal state (transforming it into $\tilde{R}$).
Finally $\tilde{R}$ is {\em reset} back to the blank ancilla state at work cost $\cEnt{R^{t}}{X^{t+1}\ldots X^{t+k} R^{t+k}}$,
 so that the generator's internal state is ready to produce the next part of the pattern.
(See also {\em physical example} in the {\em Technical Appendix}.)
}
\end{figure}

The remaining contribution $W^k_{\rm diss}$ corresponds to the cost of updating the internal memory from $R^t$ to $R^{t+k}$ so that the generator maintains prescience. This is bounded by
\begin{align}
\beta W^k_{\rm diss}(R) & \geq \cEnt{R^{t}}{X^{t+1}\ldots X^{t+k} R^{t+k}} \nonumber \\
& \quad - \cEnt{R^{t+k}}{R^t X^{t+1}\ldots X^{t+k}}, \label{eq:cEntExplicit}
\end{align}
which is always non-negative for all $k$, and all choices of memory $R$ (see lemma~\cref{lem:NonDecreasingCrypt}).
\Cref{fig:UpdateState} illustrates one possible realization of this limit, but we stress that the bound given in \cref{eq:GenerateCost} is general for any method of prescient pattern generation.
In particular, any non-degeneracy within the internal memory's Hamiltonian does not play a role in this quantity, since the process is assumed to be stationary: the average memory state remains the same at each time-step and so the memory's average energy does not change from step to step.

The first term in Eq. (\ref{eq:cEntExplicit}) represents the cost of erasing the previous state $R^t$, offset by the mutual information between the new state of the memory $R^{t+k}$ and the patterned outputs $X^{t+1}\ldots X^{t+k}$.

The second term reflects the effect of (non-)unifilarity.
Memory $R$ is defined as {\em unifilar} (see, e.g.\ \cite{MahoneyEJC11}) if $\cEnt{R^{t+k}}{R^t X^{t+1}\ldots X^{t+k}} = 0$.
If the update is not unifilar\footnote{
Causal states are automatically unifilar~\cite{ShaliziC01},
 but we do not need to make this assumption for the broader class of memory considered in this article.
 },
 according to Landauer's principle we can recover a portion of the work cost associated with the memory's change in entropy.
However, the first term in equation~\eqref{eq:cEntExplicit} may be expanded as
 $\cEnt{R^{t}}{X^{t+1}\ldots X^{t+k} R^{t+k}} = \cEnt{X^{t+1}\!\ldots\!X^{t+k}}{R^t} - \cEnt{X^{t+1}\!\ldots\!X^{t+k}}{R^{t+k}} + \cEnt{R^{t+k}}{X^{t+1}\ldots X^{t+k} R^{t}}$,
allowing us to alternatively express the bound on dissipated work as
\begin{align}
\beta W^k_{\rm diss}(R)
&  \geq   && \hspace{-2em}  \cEnt{X^{t+1}\!\ldots\!X^{t+k}}{R^t} \nonumber \\
& &&  \hspace{-2em} - \cEnt{X^{t+1}\!\ldots\!X^{t+k}}{R^{t+k}}, \label{eq:cEntNost}
\end{align}
[see lemma~\cref{thm:XrForm}(\ref{lem:part:PredRetH}) in the {\em Technical Appendix}\,].
Here, it can be seen that the non-unifilar term has been explicitly cancelled.
Whatever work might have been gained by introducing randomness into $R^{t+k}$ is entirely cancelled out by the cost of resetting this randomness in the previous state $R^t$.
Hence unifilarity does not {\em per se} play a role in deciding the thermodynamic advantage of prescient memory.

Rather, we see that the dissipative cost is proportional to the difference between $R^t$'s predictive power to guess the next $k$ symbols, and $R^{t+k}$'s {\em retrodictive} power to remember the preceding $k$ symbols.
Failure to predict increases the first entropy, and hence the amount of dissipation.
On the other hand, failure to retrodict increases the second entropy and lessens the total dissipation.

\vspace{0.5em}
\inlineheading{Extracting work from a pattern.}
Let us now evaluate how much work we can extract by consuming a pattern,
 by considering the prescient {\em extractor}.
Recall that the extractor takes $k$ systems on the tape from a configuration according to the pattern $X^{t+1}\ldots X^{t+k}$ into the default configuration $X_{\rm dflt}\ldots X_{\rm dflt}$, and must update its internal memory from $R^t$ to $R^{t+k}$.
In order to fully account for all the changes in entropy in our system,
 again we consider both the tape on which the pattern is written and the internal memory of the extractor.
In the technical appendix (see Theorem~\cref{thm:extract}), we prove that the maximum work-output when the tape Hamiltonian is \mbox{degenerate} is bounded by
\begin{equation}
\label{eq:TapeFreeEnergy}
\beta W^k_{\rm out} \leq k \left[ \Ent{X_{\rm dflt}} - \cEnt{X^{t+1}}{S^t}   \right].
\end{equation}

The work output is entirely proportional to the change in entropy of the $k$ symbols on the tape,
and has no dependence on the choice of internal memory $R$. 
If the tape Hamiltonian is not degenerate, such as would typically be the case in experimental applications, the above acquires the additional term $-\Delta E$, corresponding to a change in the expectation value of the tape's energy that is equal and opposite to the term in generation.
Unlike with the generator, it does not matter what sort of memory is used for extraction.
Here the cost of updating the memory appears to be zero.

This curiosity may be explained by carefully considering the operational difference between generators and extractors.
Firstly, note that these two processes are not exactly the reverse of each other in terms of initial and final states: in both generator and extractors, the memory advances in the same direction from $R^t$ to $R^{t+k}$ in each cycle.
Next, observe that at the end of the extraction, there is only one copy of the pattern's relevant past information (encoded in the extractor's internal memory), whereas in generation, this information is retained in both the generator's internal memory and on the tape.
The extractor can move the information from the systems on the tape $X^{t+1}\ldots  X^{t+k}$ into its internal memory $R^{t+k}$,
 but the generator must copy this information. Moving information is a logically reversible process, whereas copying information is not---and it is {\em logical irreversibility} that lies behind the dissipative costs in computation~\cite{Bennett82}.
This subtle, but important, distinction reveals to us why updating the memory must dissipate heat during pattern generation, but not pattern extraction.

\vspace{0.5em}
\inlineheading{Simpler is thermodynamically better}.
We are now in a position to consider the whole thermodynamic cycle, as illustrated in \cref{fig:SimExtCycle}.
What is the minimal amount of heat dissipation in such a cyclic process?
Suppose a generator and extractor make use of a particular encoding map resulting in prescient memory governed by $R$.
In in limit where both devices are implemented optimally at the same inverse temperature $\beta$, such that the inequalities~\eqref{eq:GenerateCost}~and~\eqref{eq:TapeFreeEnergy} are both saturated, we find that $W^k_{\rm gen}(R) - W^k_{\rm out} = \mathcal{W}^k_{\rm diss} (R) $,
where (as per ineq.~\eqref{eq:cEntNost}):
\begin{align}
 \beta \mathcal{W}^k_{\rm diss} (R)&= \cEnt{X^{t+1}\!\ldots\!X^{t+k}}{R^t} \nonumber \\
&  - \cEnt{X^{t+1}\!\ldots\!X^{t+k}}{R^{t+k}}, \label{eq:MinimumDissipationEq}
\end{align}
Physically, this represents the minimal amount of heat we must dissipate given a particular choice of prescient memory, given that we generate $k$ steps of the pattern per cycle. We now state two immediate results:

\vspace*{0.5em}

\mainresult{res:CrypticDissipation}{
$W^k_{\rm diss} > 0$ whenever $\Info{R}{\past{X}} >  \Info{\past{X}}{\future{X}}$: If our choice of prescient states stores more information about the past of a pattern than the information that the past contains about the future, then {\em any} thermodynamic cycle based on such states will be wasteful.
}
\vspace*{0.5em}

\mainresult{res:Simpler}{
Out of all the possible prescient states $R$ to use as our internal memory, the simplest ones -- corresponding to the causal states $S$ -- dissipate the least heat, i.e., minimize $\mathcal{W}^k_{\rm diss}(R)$.
Thus simpler is thermodynamically better.
}

\vspace*{0.5em}

We briefly outline the reasoning here, leaving the formal proofs for the {\em Technical Appendix}.
Consider equation~\eqref{eq:cEntNost}.
The first term is the same for {\em all} choices of prescient memory, since they all contain equal capacity to reduce uncertainty about future outputs.
However, the second term takes its largest value when $R^{t+k}$ contains the least information about the preceding $k$ symbols. This suggests the the less information we retain about the past, the less heat is dissipated. 
The minimal dissipation is then satisfied by using the causal states.

\vspace{0.5em}\inlineheading{Thermodynamics and complexity.}
Na\"ively, one may expect that it is always possible (in principle) to set  $\mathcal{W}^k_{\rm diss}(R) = 0$ by choosing prescient states such that $\Info{R}{\past{X}} = \Info{\past{X}}{\future{X}}$.
That is, given that the past of a pattern contains $b$ bits about its future, we can build a fully reversible cycle of pattern generation and extraction by ensuring our pattern manipulators' memories contain no more than $b$ bits about the pattern's past.

Surprisingly, a fundamental result in computational mechanics implies this is not true. 
Given a generic pattern $\Pr{\past{X},\future{X}}$, its statistical complexity $C_\mu = \Ent{S} = \Info{\past{X}}{S}$ is generally strictly greater than the amount of information its past contains about it future, $\Info{\past{X}}{\future{X}}$.
That is, for most patterns, even the most-efficient prescient memory -- the causal states -- contain superfluous information about the past~\cite{CrutchfieldEM09,WiesnerGRV12}. 
For these patterns, $\Info{R}{\past{X}} > \Info{\past{X}}{\future{X}}$ for all choices of prescient memory $R$.
This yields us our third result:

\vspace*{0.25em}
\mainresult{res:BarbedArrow}{
A thermodynamically reversible cycle of pattern generation and extraction is impossible for any pattern where $C_\mu > \Info{\past{X}}{\future{X}}$.
}
\vspace*{0.25em}

In {\em Technical Appendix} (see lemma~\cref{thm:XrForm}(\ref{lem:part:RetPredI})), we also prove that the minimum dissipative cost may be expressed as a difference in mutual information
\begin{align}
\beta \mathcal{W}^k_{\rm diss}(R) & = \Info{X^{t+1}\!\ldots\!X^{t+k}}{R^{t+k}} \nonumber\\
& \quad - \Info{X^{t+1}\!\ldots\!X^{t+k}}{R^t}.  \label{eq:InfoNost}
\end{align}
In the instantaneous limit $k=1$, this recovers a mathematical quantity similar to that introduced by Still et al.~\cite{StillSBC12} as the {\em useless instantaneous nostalgia}.
This coincidence is particularly striking, given that the two results are derived using very different mathematical frameworks.

Increasing the number of steps of the pattern processed per cycle $k$ never increases the amount of dissipation {\em per step of the pattern} (theorem~\cref{thm:ThermoCausality}).
(Cf.\ the dissipation {\em per cycle}, which certainly must not decrease with the step size -- theorem~\cref{thm:NonDecreasePhysical}.)
In the limit of large $k$, the minimum dissipation of produce a string of $k$ parts of the pattern tends towards $\Info{R}{\past{X}} - \Info{\past{X}}{\future{X}}$.
Any attempts using smaller $k$ to produce the same number amount of the pattern will dissipate at least as much work as this.
Moreover, when most efficient memory -- the causal states $S$ -- are used, this limiting quantity becomes the {\em crypticity} of a process $\chi = C_\mu - \Info{\past{X}}{\future{X}}$, which captures the minimal amount of superfluous information that {\em any} predicative model of the given pattern must unavoidable store~\cite{CrutchfieldEM09,MahoneyEC09,MahoneyEJC11,JamesMEC14}, coinciding with a previous result of Wiesner et al.~\cite{WiesnerGRV12}.

We hope that future work will also bridge the results of this article with specific physical frameworks, such as presented in \cite{MandalJ12}.
This requires further development in the nascent field of continuous time computational \mbox{mechanics}~\cite{MarzenC16}, 
 to identify whether the process in~\cite{MandalJ12} is indeed an example of prescient pattern manipulator (as presented in this text), 
 or if its internal state instead describes another non-prescient type of memory (e.g.\ containing oracular information~\cite{CrutchfieldEJM10}).

\begin{table*}[hbt]
\renewcommand{\arraystretch}{1.55}
\begin{tabular}{ | c | c | c | c | c | c |}
\hline
Machine & Tape cost $\beta W^k_{\rm tape}$ & Extractable work $\beta W^k_{\rm out}$ & Dissipation $\beta W^k_{\rm diss}$
& Dissipation per output
\\ 
name & Ineq.~\eqref{eq:TapeCost} & Ineq.~\eqref{eq:TapeFreeEnergy} & Ineqs.~\eqref{eq:cEntExplicit}, \eqref{eq:cEntNost}, \eqref{eq:InfoNost} or \eqref{eq:MinimumDissipationEq}.
& $\beta W^k_{\rm diss} / k$
\\
\hline
$\epsilon-$machine ($k=1$)  &
 $\left[ \Ent{X_{\rm dflt}} - h\!\left(p\right) \right]$ &
 $\left[ \Ent{X_{\rm dflt}} - h\!\left(p\right) \right]$ &
 $h\!\left(p\right)$ &
 $h\!\left(p\right)$
\\
$\epsilon-$machine ($k=2$)  &
 $2\left[ \Ent{X_{\rm dflt}} - h\!\left(p\right) \right]$ &
 $2\left[ \Ent{X_{\rm dflt}} - h\!\left(p\right) \right]$ &
 $h\!\left(p\right)$ &
 $\dfrac{h\!\left(p\right)}{2}$
\\
$\epsilon-$machine ($k\geq3$)  &
 $k\left[ \Ent{X_{\rm dflt}} - h\!\left(p\right) \right]$ &
 $k\left[ \Ent{X_{\rm dflt}} - h\!\left(p\right) \right]$ &
 $h\!\left(p\right)$ &
 $\dfrac{h\!\left(p\right)}{k}$
\\
\hline
``Last two'' ($k=1$)  &
 $\left[ \Ent{X_{\rm dflt}} - h\!\left(p\right) \right]$ &
 $\left[ \Ent{X_{\rm dflt}} - h\!\left(p\right) \right]$ &
 $h\!\left(p\right)$ &
 $h\!\left(p\right)$
\\
``Last two'' ($k=2$)  &
 $2\left[ \Ent{X_{\rm dflt}} - h\!\left(p\right) \right]$ &
 $2\left[ \Ent{X_{\rm dflt}} - h\!\left(p\right) \right]$ &
 $2 h\!\left(p\right)$ &
 $h\!\left(p\right)$
\\
``Last two'' ($k\geq3$)  &
 $k\left[ \Ent{X_{\rm dflt}} - h\!\left(p\right) \right]$ &
 $k\left[ \Ent{X_{\rm dflt}} - h\!\left(p\right) \right]$ &
 $2 h\!\left(p\right)$ &
 $\dfrac{2}{k} h\!\left(p\right)$
\\
\hline
``Last $N$'' ($k\leq N$)  &
 $k\left[ \Ent{X_{\rm dflt}} - h\!\left(p\right) \right]$ &
 $k\left[ \Ent{X_{\rm dflt}} - h\!\left(p\right) \right]$ &
 $k h\!\left(p\right)$ &
 $h\!\left(p\right)$
\\
``Last $N$'' ($k >  N$)  &
 $k\left[ \Ent{X_{\rm dflt}} - h\!\left(p\right) \right]$ &
 $k\left[ \Ent{X_{\rm dflt}} - h\!\left(p\right) \right]$ &
 $N h\!\left(p\right)$ &
 $\dfrac{N}{k} h\!\left(p\right)$
\\
\hline
\end{tabular}
\caption{
\label{table:PerturbedCoinCosts}
\caphead{Minimum thermodynamic costs for perturbed coin process.}
Here $h\!\left(p\right) := -p\log p - \left(1\!-\!p\right) \log\left(1\!-\!p\right)$ denotes the binary entropy. To arrive at the value in these columns, we calculate that $\cEnt{X^{t+1}}{S^t} = h\!\left(p\right)$ by noting that both $H$ and $T$ states are equally likely to occur, and both have an uncertainty of $h\!\left(p\right)$ as to the value of the next output.
It can be seen immediately from the form of  eq.~\eqref{eq:TapeCost} and eq.~\eqref{eq:TapeFreeEnergy} that these values scale linearly with $k$: the number of pattern elements written or consumed per cycle.
}
\renewcommand{\arraystretch}{1}
\end{table*}

\inlineheading{General framework of pattern manipulation.}
The two types of device---generators and extractors---that we have presented in this article are building blocks that can be combined into complex pattern manipulators.
The simplest example is the cycle of generation followed by extraction at the same temperature (\cref{fig:SimExtCycle}).
This configuration could be considered as charging a battery that stores energy in the form of a pattern, to be later released by the extractor.
Since the contributions from writing and consuming the pattern cancel out, the net cost is from the work dissipation whilst updating the generator's memory.
Using the simplest internal memory ensures that the least work is wasted.

However, the cycle presented in \cref{fig:SimExtCycle} could be modified such that the generator and extractor act at {\em different} temperatures.
In the scenario where the extractor bath temperature $T_E$ is greater than the generator bath temperature $T_G$ and $\cEnt{X^{t+1}}{S^t} < \Ent{X_{\rm dflt}}$,
 and the tape is subject to a degenerate Hamiltonian, the system will function as a {\em heat-engine} with efficiency $\eta$ given:
\begin{align}
\label{eq:CarnotEff}
\eta = & 1 - \dfrac{T_G}{T_E} - \dfrac{T_G \chi_R(k)}{T_E k \left[ \Ent{X_{\rm dflt}} - \cEnt{X^{t+1}}{S^t}  \right]}.
\end{align}
where $\chi_R(k) = \cEnt{R^{t}}{X^{t+1}\ldots X^{t+k} R^{t+k}} - \cEnt{R^{t+k}}{R^t X^{t+1}\ldots X^{t+k}}$
 is the right hand side of equation~\eqref{eq:MinimumDissipationEq}. 
 It follows that the Carnot efficiency $\eta_C = 1 - \frac{T_E}{T_G}$ can only be achieved when the statistical complexity $C_\mu = \Info{\past{X}}{\future{X}}$ such that the pattern has zero crypticity.

\vspace*{0.5em}
\inlineheading{Simple Example.}
Let us explicitly evaluate the thermodynamics of a simple pattern described by the {\em perturbed coin} process.
Envision a coin in a box that takes one of two possible configurations: $\{x\} = \{ H, T\}$ (standing for ``heads'' and ``tails'' respectively).
At each discrete time step, the box is perturbed such that with probability $p\in(0,1)$ the coin flips from $H$ to $T$ (or vice versa).
The pattern $\pastfuture{X}$ consists of bi-infinite string of random variables describing which side of the coin faces up at each time step.

It is clear that this pattern is Markovian: the statistics of future outputs $\future{X}$ depend only on the very last output $X^{t}$, corresponding to state of coin at the current time-step. 
The past thus divides into two causal states, denoted $s_H$ and $s_T$, corresponding to the two possible values of $X^{t}$. 
The statistics of the pattern can then be represented by transitions between these causal states (see \cref{fig:PerturbedCoin}).
By symmetry, $\Pr{S=s_H} = \Pr{S=s_T} = \frac{1}{2}$, and so the process has {\em statistical complexity} $C_\mu = 1\;\mathrm{bit}$.

\begin{figure}[hbt]
\includegraphics[width=0.375\textwidth]{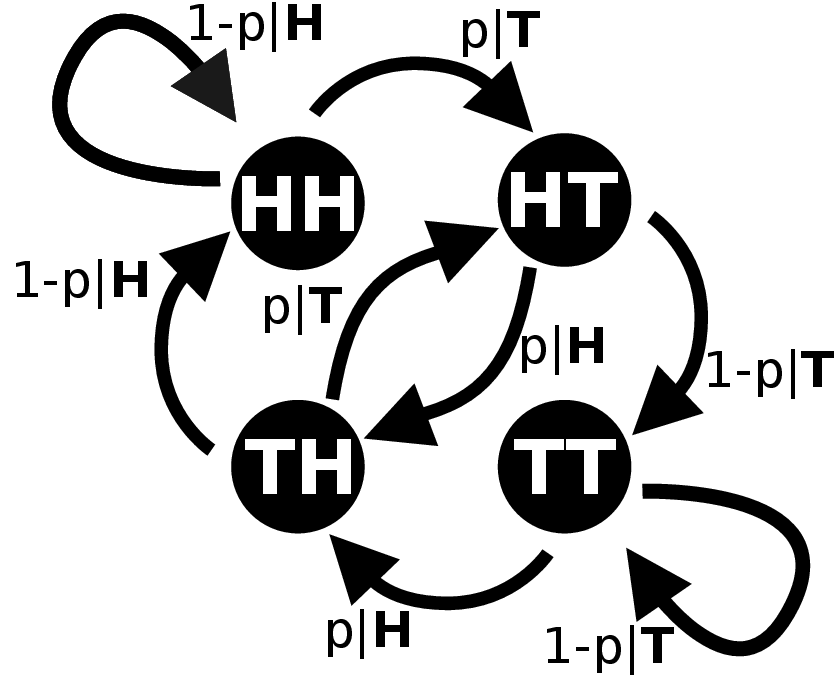}
\caption{
\label{fig:PerturbedCoinLastTwo}
\caphead{Perturbed coin process ``Last two'' machine ($k=1$).}
The random walk on this network produces exactly the same pattern as the $\epsilon-$machine (\cref{fig:PerturbedCoin}). This machine's memory corresponds to the last two outcomes of the pattern.
}
\end{figure}

Here we study the thermodynamic quantities when running a thermodynamic cycle at different strides $k$ using a variety of different machines:
 (1)~The simplest pattern manipulators, that stores only the causal states, 
 (2)~the ``last two'' machines that stores the last two values the coin took (i.e., setting $R = X^{t-1}X^{t}$ as per \cref{fig:PerturbedCoinLastTwo}), 
 and (3)~a more general ``last $N$'' machine that stores the last $N$ outputs (i.e., setting $R = X^{t-n+1}\ldots X^{t-1}X^{t}$). 
 The cases $N=1$ and $N=2$ correspond to the $\epsilon-$machine and ``last two'' machine respectively.

The change in free energy for $k$-symbols of the pattern [eq.~\eqref{eq:TapeCost}], extractable work when consuming $k$-symbols of the pattern [eq.~\eqref{eq:TapeFreeEnergy}], and unavoidable heat dissipation are summarized in \cref{table:PerturbedCoinCosts}.
Detailed calculations can be found in the {\em Technical Appendix}. 
We make several observations:
\begin{enumerate}[leftmargin=1em]
\item The free energy change of the tape from eq.~\eqref{eq:TapeCost} and the extractable work from eq.~\eqref{eq:TapeFreeEnergy} are always equal and opposite. 
Although this cost must be accounted for when running the generator or extractor in isolation, the two cancel in a thermodynamic cycle.
\item Choosing different $X_{\rm dflt}$ affects the amount of work required to write to the tape, and the amount of work generated when the tape is consumed, but does not change the total heat dissipated per cycle. 
Suppose that $X_{\rm dflt}$ is maximally mixed, such that $\Ent{X_{\rm dflt}} = 1\;\rm{bit}$. 
In this case, work must be invested to write the pattern to the tape (except when $p=\frac{1}{2}$). 
If instead $X_{\rm dflt}$ is always heads, such that $\Ent{X_{\rm dflt}}= 0$, then we would instead extract $h\!\left(p\right):= -p\log p - \left(1\!-\!p\right) \log\left(1\!-\!p\right)$ units of work from each time-step (at the cost of reducing the free energy of the tape). 
Another interesting choice would be to set $\Ent{X_{\rm dflt}} = h\!\left(p\right)$, such that the free energy of the tape does not change during generation or consumption, and only the dissipative terms remain. 
In all cases, the total amount of heat dissipated over the entire cycle is the same.
\item Simpler is better.
Out of all possible last-$N$ machines, the one that minimizes the dissipation per cycle occurs when $N=1$, corresponding to the simplest machines that store only the current causal state of the past. 
Heat dissipation is monotonically non-decreasing with $N$. 
Note, however, there are subtleties:
For a given stride $k$, the penalty for remembering more of the past `saturates' when $N = k$. 
This is because when the stride $k < N$, part of the memory can be updated in a logically reversible way (see appendix for details).
\item If one wants to create as much of the pattern as possible with minimal heat dissipation per output symbol, it is generally advantageous to use large strides. 
For example, when using $\epsilon$-machines, the heat dissipation remains fixed at $h(p)$ for all $k$. 
Thus that generating $k$ outputs per cycle will reduce the dissipation per symbol by a factor of $k$.
\end{enumerate}
Finally, the ultimate limits on heat dissipation can be immediately evaluated by noting that the crypticity of the pattern is non-zero. 
In particular the pattern has crypticity $\chi = C_{\mu} - I(\past{X},\future{X}) = h(p)$. 
This bounds the minimal dissipation regardless of stride, or what machine is used. 

In the {\em Technical Appendix}, we provide a discussion of this pattern manipulation, as it could be implemented in a more immediately physical framework.
There, we provide one mechanism for implementing pattern manipulations at work costs that saturate the bounds of the first row of \cref{table:PerturbedCoinCosts}, and consider explicitly the efficiency of a heat engine constructed from a cycle of generation followed by extraction.

\enlargethispage{\baselineskip}
\inlineheading{Discussion and outlook.}
Here we have presented a formal framework for treating patterns as thermodynamic resources that require energy to synthesize, which can then in turn be used to perform work. 
This involved the study of generators that convert work into a pattern, and extractors that consume patterns to perform useful work.
These components were then combined to construct a full thermodynamic cycle of pattern generation and consumption. 
We then identified the dissipative work costs involved within such a cycle, and related it to the complexity of the generators and extractors used. 
We found that {\em simpler is thermodynamically better}; the less past information retained, the less dissipation. 
Optimality is achieved when we retain just enough information to be able to replicate desired operational behaviour. 
The resultant unavoidable dissipation is then given by the crypticity of the pattern. 
These relations present thermodynamic interpretations for a fundamental complexity-theoretic property.

One can extend this framework to consider alternative scenarios where generators and extractors act in parallel on different patterns.
For example, consider a scenario where an extractor consumes one pattern, and the energy released is then used to power a generator that writes a different pattern. 
This could approximate the actions of a living organism: for example, a lion metabolizes the structure of an antelope (destroying it in the process), and uses the energy released to build more lion. 
Using the simplest internal memory for generation grants the advantage that less antelope needs to be consumed in order to produce the same amount of lion.

A foundational question of interest is whether this unavoidable heat dissipation is truly fundamental, or could it somehow be surpassed with more exotic information processing. 
Recent research indicates that quantum processors can generate certain statistical patterns of behaviour more simply~\cite{GuWRV12}, resulting in a push to generalize computational mechanics into the quantum regime~\cite{SuenTGVG15,MahoneyAC16,ThompsonGVG16,AghamohammadiMC16,RiechersMAC16}.
Could this simplicity also yield thermodynamic advantage? Any such results would present exciting new thermodynamic signatures of quantumness.

\section*{Acknowledgements}
We are grateful for funding from the John Templeton Foundation Grant 53914 {\em ``Occam's Quantum Mechanical Razor: Can Quantum theory admit the Simplest Understanding of Reality?''};
 the Foundational Questions Institute (``Physics of the Obserer'' large grant);
 the National Research Foundation, Prime Minister’s Office, Singapore, under its Competitive Research Programme (CRP Award No. NRF-CRP14-2014-02); 
 the National Research Foundation NRF-Fellowship (NRF-NRFF2016-02);
 the Ministry of Education in Singapore, the Academic Research Fund Tier 3 MOE2012-T3-1-009;
 the National Basic Research Program of China Grants 2011CBA00300 and 2011CBA00302;
 the National Natural Science Foundation of China Grants 11450110058, 61033001 and 61361136003;
 the 1000 Talents Program of China;
 and the Oxford Martin School.
We thank
 \mbox{Blake Pollard},
 \mbox{Merlin Seller},
 and \mbox{Whei-Yeap Suen}
 for illuminating \mbox{discussion}.

\balancecolsandclearpage

\newpage

\appendix
\section*{TECHNICAL APPENDIX}

\begin{lemma}[Cycle remains in step]
\label{lem:InStep}
Consider a generator with prescient memory $R$ and an extractor with prescient memory $R'$.
At time $t$, let these memories initially be in configurations $r$ and $r'$ respectively, satisfying
 $\cPr{\future{X}^t}{R=r} = \cPr{\future{X}^t}{R=r'}$.
After $k$ symbols are produced by the generator and then subsequently consumed by the extractor,
 the two devices remain ``in step'' such that
$\cPr{\future{X}^{t+k}}{R=\tilde{r}} = \cPr{\future{X}^{t+k}}{R=\tilde{r}'}$
 for the updated configurations $\tilde{r}$ and $\tilde{r}'$.

\begin{proof}
This follows almost immediately from the fact that both devices maintain the prescience of their memory.
When a prescient generator produces $k$ steps of the pattern $X^{t+1}\ldots X^{t+k} = x^{t+1}\ldots x^{t+k}$,
 it updates its memory to a new prescient state with configuration $\tilde{R} = \tilde{r}$ that satisfies
\begin{align}
\cPr{\future{X}^{t+k}}{\tilde{R} = \tilde{r}} \hspace{-5em} & \nonumber \\
& = \cPr{\future{X}^{t+k}}{R=r, X^{t+1}\ldots X^{t+k} = x^{t+1}\ldots x^{t+k}}.
\label{eq:GenInStepUR}
\end{align}
The generator can only produce sequences $x^{t+1}\ldots x^{t+k}$ that satisfy $\cPr{X^{t+1}\ldots X^{t+k}\!=\!x^{t+1}\ldots x^{t+k}}{R=r}>0$.
Similarly, when the prescient extractor receives a sequence,
 it also updates its internal memory to some state $\tilde{R}' = \tilde{r}'$ such that
\begin{align}
\cPr{\future{X}^{t+k}}{\tilde{R}' = \tilde{r}'} \hspace{-5em} & \nonumber \\
& = \cPr{\future{X}^{t+k}}{R'=r', X^{t+1}\ldots X^{t+k} = x^{t+1}\ldots x^{t+k}}.
\label{eq:ExInStepUR}
\end{align}

We now show that if the devices are initially in step, such that $\cPr{\future{X}^{t}}{R=r} = \cPr{\future{X}^{t}}{R'=r'}$,
 the above updates involving the same string will keep them in step.

The semi-infinite future of a pattern expected at the initial time $t$, $\future{X}^t$, can be split into a finite string of length $k$, $X^{t+1}\ldots X^{t+k}$, and another semi-infinite string $\future{X}^{t+k} = X^{t+k+1}\ldots$.
Noting that both $r$ and $r'$ initially predict the same statistics for $\future{X}^t$ (and hence also for any finite substring thereof),
 for any realization of the future $\future{x}^t = x^{t+1}\ldots x^{t+k} \future{x}^{t+k}$,
it follows from the probability chain rule on $\future{X}^t$ conditioned on the initial state of both devices:
\begin{align}
\cPr{\future{X}^t\!=\!\future{x}^t }{R=r} = \cPr{\future{X}^{t}=\future{x}^{t} }{R'=r'} \hspace{-18em} & \nonumber \\
& = \cPr{X^{t+1}\ldots X^{t+k}\!=\!x^{t+1}\ldots x^{t+k} }{R=r} \nonumber\\
& \quad\cdot \cPr{\future{X}^{t+k}=\future{x}^{t+k}}{R=r, X^{t+1}\ldots X^{t+k} = x^{t+1}\ldots x^{t+k}}, \nonumber \\
& = \cPr{X^{t+1}\ldots X^{t+k} = x^{t+1}\ldots x^{t+k} }{R'=r'} \nonumber \\
& \quad\cdot \cPr{\future{X}^{t+k}=\future{x}^{t+k}}{R'=r', X^{t+1}\ldots X^{t+k} = x^{t+1}\ldots x^{t+k}},
\end{align}
and since also
\begin{align}
\cPr{X^{t+1}\ldots X^{t+k}\!=\!x^{t+1}\ldots x^{t+k} }{R=r} \hspace{-15em} & \nonumber\\
& =\cPr{X^{t+1}\ldots X^{t+k}\!=\!x^{t+1}\ldots x^{t+k} }{R'=r'},
\end{align}
we can then conclude that
\begin{align}
\cPr{\future{X}^{t+k}\!=\!\future{x}^{t+k}}{R\!=\!r, X^{t+1}\ldots X^{t+k}\!=\!x^{t+1}\ldots x^{t+k}} \hspace{-22em} & \nonumber \\
& = \cPr{\future{X}^{t+k}\!=\!\future{x}^{t+k}}{R'\!=\!r', X^{t+1}\ldots X^{t+k}\!=\!x^{t+1}\ldots x^{t+k}},
\end{align}
for arbitrary $x^{t+1}\ldots x^{t+k}$, and all possible futures $x^{t+k}$.
Thus, substituting in equations~\eqref{eq:GenInStepUR} and \eqref{eq:ExInStepUR}:
\begin{equation}
\cPr{\future{X}^{t+k}}{\tilde{R} = \tilde{r}} = \cPr{\future{X}^{t+k}}{\tilde{R}' = \tilde{r}'}.
\end{equation}
The devices hence remain in step after each cycle.

\end{proof}
\end{lemma}

\begin{theorem}[Work cost of generation]
\label{thm:generate}
For a generator of a pattern $\pastfuture{X}$ that maintains prescient memory $R$,
 the work cost $W^k_{\rm gen}$ to configure $k$ systems according to the pattern
 is bounded by:
\begin{align}
\beta W^k_{\rm gen} (R) \geq~ & k \left[ \Ent{X_{\rm dflt}} - \cEnt{X^{t+1}}{S^{t}} \right] \nonumber \\
& + \cEnt{R^{t}}{X^{t+1}\ldots X^{t+k} R^{t+k}} \nonumber \\
& \quad - \cEnt{R^{t+k}}{R^t X^{t+1}\ldots X^{t+k}}.
\label{eq:thmGenAppendix}
\end{align}
\begin{proof}
We prove this using information theoretical means.
In the generation process, the generator's internal memory is initially distributed according to $R^t$, and the $k$ systems on the tape according to $X_{\rm dflt}\ldots X_{\rm dflt}$.
After generation, the final state of the generator's internal memory is $R^{t+k}$, and the $k$ systems on the tape are in states $X^{t+1}\ldots X^{t+k}$.

As such, a device-independent lower bound on the cost of generation may be found by considering Landauer's principle as applied to these states:
\begin{equation}
\beta W_{\rm gen}(R) \geq \Ent{R^t X_{\rm dflt}\ldots X_{\rm dflt}} - \Ent{R^{t+k}X^{t+1}\ldots X^{t+k}}.
\end{equation}

First, we note that $\Ent{R^t X^{t+1} \ldots X^{t+k} R^{t+k}}$ may be expanded using the chain rule in two different ways:
\begin{align}
\Ent{R^t X^{t+1}\ldots X^{t+k} R^{t+k}}  \hspace{-6em} & \nonumber \\
& = \Ent{R^{t}} + \cEnt{X^{t+1}}{R^t} + \ldots \nonumber \\
& \quad + \cEnt{X^{t+k}}{R^{t}X^{t+1}\ldots X^{t+k-1}} \nonumber \\
& \quad + \cEnt{R^{t+k}}{R^{t}X^{t+1}\ldots X^{t+k}}  \nonumber \\
& = \Ent{R^{t+k}} + \cEnt{X^{t+k}}{R^{t+k}} + \ldots \nonumber \\
& \quad + \cEnt{X^{t+1}}{R^{t+k}X^{t+k}\ldots X^{t+2}} \nonumber \\
& \quad + \cEnt{R^{t}}{R^{t+k}X^{t+k}\ldots X^{t+1}}.
\end{align}
This allows us to re-express the final state entropy $\Ent{R^{t+k}X^{t+1}\ldots X^{t+k}}$ in the form
\begin{align}
\Ent{R^{t+k}X^{t+1} \ldots X^{t+k}} \hspace{-8em} \nonumber \\
& = \Ent{R^{t+k}} + \cEnt{X^{t+k}}{R^{t+k}} + \ldots \nonumber \\
& \quad + \cEnt{X^{t+1}}{R^{t+k}X^{t+k}\ldots X^{t+2}}, \nonumber\\
& = \Ent{R^t} + \cEnt{X^{t+1}}{R^t} + \ldots  \nonumber \\
& \quad + \cEnt{X^{t+k}}{R^t X^{t+1}\ldots X^{t+k-1}} \nonumber \\
& \quad + \cEnt{R^{t+k}}{R^{t}X^{t+1}\ldots X^{t+k}} \nonumber \\
& \quad\quad - \cEnt{R^{t}}{R^{t+k}X^{t+k}\ldots X^{t+1}}. \label{eq:GenFinalStateEntropy}
\end{align}

Moreover, since the internal memory $R$ is prescient about the future of $X$,
 and all choices of prescient memory must correspond to a fine-graining of the causal states $S^{t}$~(see lemma~7 in~\cite{ShaliziC01}).
As such $R^t X^{t+1} \ldots X^{t+j}$ is sufficient to perfectly determine $S^{t+j}$,
 which is in turn as exactly as useful as $R^t X^{t+1} \ldots X^{t+j}$ for predicting values of $X^{t+j+1}$ onwards.
Thus the term $\cEnt{X^{t+j}}{S^{t+j-1}}$ may be substituted for every addend of the form $\cEnt{X^{t+j}}{R^t X^{t+1} \ldots X^{t+j-1}}$ in equation~\eqref{eq:GenFinalStateEntropy}.
Moreover, by stationarity, $\cEnt{X^{t+1}}{S^{t}} = \cEnt{X^{t+2}}{S^{t+1}} = \ldots$, and hence the sum over $k$ of these terms may be combined into the single expression $k\cEnt{X^{t+1}}{S^t}$.
As such, we can re-express the final entropy following generation as
\begin{align}
\Ent{R^{t+k}X^{t+1} \ldots X^{t+k}} = \hspace{-6em}& \nonumber \\
& \Ent{R^t} +  k\cEnt{X^{t+1}}{S^t}\nonumber \\
& + \cEnt{R^{t+k}}{R^{t}X^{t+1}\ldots X^{t+k}} \nonumber \\
& \quad - \cEnt{R^{t}}{R^{t+k}X^{t+k}\ldots X^{t+1}}. \label{eq:finalEntIntermed}
\end{align}

Meanwhile, the initial entropy $\Ent{R^t X_{\rm dflt}\ldots X_{\rm dflt}}$ can straightfowardly be written
\begin{equation}
\Ent{R^t X_{\rm dflt}\ldots X_{\rm dflt}} = \Ent{R^t} + k\Ent{X_{\rm dflt}},
\end{equation}
since there are no correlations within the initial state.

Hence, from Landauer's principle, the difference between these two quantities yields the minimum work exchange required to generate a pattern and update the generator's internal memory:
\begin{align}
\label{eq:WorkCostAppPf}
\beta W^k_{\rm gen}(R) \geq~ & k \left[ \Ent{X_{\rm dflt}} - \cEnt{X^{t+1}}{S^{t}} \right] \nonumber \\
& + \cEnt{R^{t}}{X^{t+1}\ldots X^{t+k} R^{t+k}} \nonumber \\
& \quad - \cEnt{R^{t+k}}{R^t X^{t+1}\ldots X^{t+k}}.
\end{align}

\end{proof}
\end{theorem}

\begin{theorem}[Work output from extraction]
\label{thm:extract}
No extractor with prescient internal memory $R$ can {\em extract} from $k$ symbols of a pattern $\pastfuture{X}$,
 more work $W^k_{\rm out}$ than the following bound:
\begin{align}
\beta W^k_{\rm out} \leq~ & k \left[ \Ent{X_{\rm dflt}} - \cEnt{X^{t+1}}{S^{t}} \right].
\label{eq:extractApp}
\end{align}
The choice of internal memory plays no limiting role in determining the amount of work extractable.
\begin{proof}
As with theorem~\cref{thm:generate}, we consider the fundamental limitations placed by Landauer's principle.
A cyclically operating prescient extractor must result in the transformation of the tape from $X^{t+1}\ldots X^{t+k}$ to $X_{\rm dflt}\ldots X_{\rm dflt}$, while updating its internal memory from $R^t$ to $R^{t+k}$.
As such, the bound from Landauer's principle is
\begin{equation}
\beta W^k_{\rm out} \leq \Ent{R^{t+k} X_{\rm dflt}\ldots X_{\rm dflt}} - \Ent{R^{t} X^{t+1} \ldots X^{t+k}}.
\end{equation}

In this case, the final entropy $\Ent{R^{t+k} X_{\rm dflt}\ldots X_{\rm dflt}}$ is easy to calculate, since there is no correlation between any of the variables:
\begin{equation}
\Ent{R^{t+k} X_{\rm dflt}\ldots X_{\rm dflt}} = \Ent{R^{t+k}} + k \Ent{X_{\rm dflt}}.
\end{equation}

On the other hand, the entropy of the initial state $\Ent{R^t X^{t+1}\ldots X^{t+k}}$
 is lower than if the variables were independent, since $R^t$ contains some information that can be used to infer $X^{t+1}\ldots X^{t+k}$.
We can use the chain rule expansion to see that this entropy is
\begin{align}
\Ent{R^t X^{t+1}\ldots X^{t+k}} = \hspace{-5em} & \nonumber \\
& \Ent{R^t} + \cEnt{X^{t+1}}{R^t} + \ldots \nonumber \\
& \quad + \cEnt{X^{t+k}}{R^t X^{t+1}\ldots X^{t+k-1}}.
\end{align}

Using exactly the same logic as in theorem~\cref{thm:generate}, each term of the form $\cEnt{X^{t+j}}{R^t X^{t+1}\ldots X^{t+j-1}}$ can be substituted with a term $\cEnt{X^{t+j}}{S^{t+j-1}}$ (because the memory is prescient), and from stationarity these terms are all the same, and so may be collectively replaced by $k\cEnt{X^{t+1}}{S^t}$.
Thus, the initial entropy is
\begin{equation}
\Ent{R^t X^{t+1}\ldots X^{t+k}} = \Ent{R^t} + k\cEnt{X^{t+1}}{S^t}.
\end{equation}

Finally, from stationarity we note that $\Ent{R^t} = \Ent{R^{t+k}}$ and so computing the difference in these entropies, and applying Landauer's principle, we arrive at the bound
\begin{align}
\beta W^k_{\rm out} \leq~ & k \left[ \Ent{X_{\rm dflt}} - \cEnt{X^{t+1}}{S^{t}} \right],
\end{align}
which has no explicit dependence on $R$.

\end{proof}
\end{theorem}

\inlineheading{Dissipative work term.}
As discussed in the article,
 since the work cost in the first line of \cref{eq:thmGenAppendix} of \cref{thm:generate}
 has no dependence on the choice of memory used, but is entirely a function of the pattern on the tape,
 and moreover is exactly equal to the energy that may be recovered according to \cref{thm:extract},
 we may naturally divide the cost into two parts:
\begin{align}
\beta W^k_{\rm tape} \geq~ & k \left[ \Ent{X_{\rm dflt}} - \cEnt{X^{t+1}}{S^{t}} \right], \label{eq:AppTapeCost} \\
\beta W^k_{\rm diss}(R) \geq~ & \cEnt{R^{t}}{X^{t+1}\ldots X^{t+k} R^{t+k}} \nonumber \\
& \quad - \cEnt{R^{t+k}}{R^t X^{t+1}\ldots X^{t+k}} \label{eq:AppDissDef}.
\end{align}

In an optimally-implemented cycle of generation and extraction of the same pattern using the same type of memory $R$ in both devices (where both run at the same temperature), the change in work is given by the quantity that saturates \cref{eq:AppDissDef}.
For notational convenience, we denotate this minimum difference as $\chi_R(k)$,
\begin{align}
\chi_R(k) & := \cEnt{R^{t}}{X^{t+1}\ldots X^{t+k} R^{t+k}} \nonumber \\
& \quad - \cEnt{R^{t+k}}{R^t X^{t+1}\ldots X^{t+k}},
\end{align}
such that theorem~\cref{thm:generate} may be rephrased as:
\begin{equation}
\beta W^k_{\rm diss}(R) \geq~ \chi_R(k).
\end{equation}

We can garner some additional physical insight about $\chi_R(k)$, by expressing it in a few alternative forms, listed below:-

\begin{lemma}[Equivalent forms of $\chi_R(k)$]
\label{thm:XrForm}
For prescient memory $R$, the following expressions are equal:

\begin{enumerate}[label=\roman*.,ref=\roman*]
\item Memory reset - non-unfilarity: \label{lem:part:RetroNonUni}
\begin{align}
\chi_R(k) & = \cEnt{R^{t}}{X^{t+1}\ldots X^{t+k} R^{t+k}} \nonumber \\
& \quad - \cEnt{R^{t+k}}{R^t X^{t+1}\ldots X^{t+k}},
\end{align}

\item Past memory - future memory uncertainty: \label{lem:part:PastFutMemH}
\begin{align}
\label{eq:MarkovDisp}
\chi_R(k) & =
\cEnt{R^t}{X^{t+1}\ldots X^{t+k}} \nonumber \\
& \quad - \cEnt{R^{t+k}}{X^{t+1}\ldots X^{t+k}}.
\end{align}

\item Predictive - retrodictive uncertainty: \label{lem:part:PredRetH}
\begin{align}
\chi_R(k) & = \cEnt{X^{t+1}\ldots X^{t+k}}{R^t} \nonumber \\
& \quad - \cEnt{X^{t+k}\ldots X^{t+1}}{R^{t+k}}.
\end{align}

\item Retrodictive - predictive information: \label{lem:part:RetPredI}
\begin{align}
\chi_R(k) & =
\Info{X^{t-k+1}\ldots X^{t}}{R^{t}} \nonumber \\
& \quad - \Info{X^{t+k}\ldots X^{t+1}}{R^{t}}.
\end{align}

\item Pattern-memory - memory-pattern block entropy: \label{lem:part:blockH}
\begin{align}
\label{eq:BlockEntropy}
\chi_R(k) & =\Ent{X^{t+1}\ldots X^{t+k}R^{t}} \nonumber \\
& \quad - \Ent{X^{t+k}\ldots X^{t+1}R^{t+k}}.
\end{align}

\item Difference in memory retrodictability: \label{lem:part:MemRetro} \label{cor:InfiniteFuture}
\begin{align}
\label{eq:BlockEntropy}
\chi_R(k) & = \cEnt{R^t}{\future{X}} - \cEnt{R^{t+k}}{\future{X}}.
\end{align}
\end{enumerate}

\begin{proof}

\ref{lem:part:RetroNonUni}.
We take the form of \ref{lem:part:RetroNonUni}  as our initial definition of $\chi_R(k)$.
\begin{align}
\chi_R(k) & := \cEnt{R^{t}}{X^{t+1}\ldots X^{t+k} R^{t+k}} \nonumber \\
& \quad - \cEnt{R^{t+k}}{R^t X^{t+1}\ldots X^{t+k}}.
\end{align}

\ref{lem:part:PastFutMemH}.
We may expand $\Ent{R^t X^{t+1} \ldots X^{t+k} R^{t+k}}$ in two ways:
\begin{align}
\Ent{R^t X^{t+1}\ldots X^{t+k} R^{t+k}}  \hspace{-8em} & \nonumber \\
& = \Ent{X^{t+1}\ldots X^{t+k}} + \cEnt{R^t}{X^{t+1}\ldots X^{t+k}}  \nonumber \\
& \quad + \cEnt{R^{t+k}}{R^{t}X^{t+1}\ldots X^{t+k}} \nonumber \\
& = \Ent{X^{t+1}\ldots X^{t+k}} + \cEnt{R^{t+k}}{X^{t+1}\ldots X^{t+k}}  \nonumber \\
& \quad + \cEnt{R^{t}}{R^{t+k}X^{t+1}\ldots X^{t+k}}.
\end{align}
Since the terms $\Ent{X^{t+1}\ldots X^{t+k}}$ cancel, we see
\begin{align}
& \cEnt{R^{t}}{R^{t+k}X^{t+1}\ldots X^{t+k}}  \nonumber \\
& \qquad - \cEnt{R^{t+k}}{R^{t}X^{t+1}\ldots X^{t+k}} \nonumber \\
 = & \cEnt{R^t}{X^{t+1}\ldots X^{t+k}} \nonumber \\
& \qquad - \cEnt{R^{t+k}}{X^{t+1}\ldots X^{t+k}}.
\end{align}
The top term is \ref{lem:part:RetroNonUni}, the bottom \ref{lem:part:PastFutMemH},
 hence these are equal.

\ref{lem:part:PredRetH}.
A  different expansion of the joint entropy is
\begin{align}
\Ent{R^t X^{t+1}\ldots X^{t+k} R^{t+k}}  \hspace{-6em} & \nonumber \\
& = \Ent{R^{t}} + \cEnt{X^{t+1}\ldots X^{t+k}}{R^t}  \nonumber \\
& \quad + \cEnt{R^{t+k}}{R^{t}X^{t+1}\ldots X^{t+k}} \nonumber \\
& = \Ent{R^{t+k}} + \cEnt{X^{t+k}\ldots X^{t+1}}{R^{t+k}} \nonumber \\
& \quad + \cEnt{R^{t}}{R^{t+k}X^{t+k}\ldots X^{t+1}}.
\end{align}
Using stationarity to set $\Ent{R^t} = \Ent{R^{t+k}}$,
\begin{align}
& \cEnt{R^{t}}{R^{t+k}X^{t+1}\ldots X^{t+k}} \nonumber \\
& \qquad - \cEnt{R^{t+k}}{R^{t}X^{t+1}\ldots X^{t+k}}\nonumber \\
= & \cEnt{X^{t+1}\ldots X^{t+k}}{R^t} - \cEnt{X^{t+k}\ldots X^{t+1}}{R^{t+k}},
\end{align}
The top term is \ref{lem:part:RetroNonUni}, the bottom is \ref{lem:part:PredRetH},
 and hence these expressions are equal.

\ref{lem:part:RetPredI}.
From the definition of mutual information $\Info{A}{B} = \Ent{A} - \cEnt{A}{B}$, and using stationarity, we can re-express $\chi_R(k)$ as
\begin{align}
\cEnt{X^{t+1}\ldots X^{t+k}}{R^t} - \cEnt{X^{t+k}\ldots X^{t+1}}{R^{t+k}} \hspace{-20em}& \nonumber \\
& = \Info{X^{t+1}\ldots X^{t+k}}{R^{t+k}} - \Info{X^{t+k}\ldots X^{t+1}}{R^{t}}, \nonumber \\
& = \Info{X^{t-k+1}\ldots X^{t}}{R^{t}} - \Info{X^{t+k}\ldots X^{t+1}}{R^{t}}.
\end{align}
The top term is \ref{lem:part:PredRetH}, the bottom is \ref{lem:part:RetPredI}
 and hence these expressions are equal.

\ref{lem:part:blockH}.
Again, we expand
\begin{align}
\Ent{R^t X^{t+1}\ldots X^{t+k} R^{t+k}}  \hspace{-10em} & \nonumber \\
& = \Ent{R^{t}X^{t+1}\ldots X^{t+k}} \nonumber \\
& \quad + \cEnt{R^{t+k}}{R^{t}X^{t+1}\ldots X^{t+k}} \nonumber \\
& = \Ent{R^{t+k}X^{t+k}\ldots X^{t+1}} \nonumber \\
& \quad + \cEnt{R^{t}}{R^{t+k}X^{t+k}\ldots X^{t+1}},
\end{align}
such that
\begin{align}
& \cEnt{R^{t}}{R^{t+k}X^{t+1}\ldots X^{t+k}} \nonumber\\
& \quad - \cEnt{R^{t+k}}{R^{t}X^{t+k}\ldots X^{t+1}} & \nonumber \\
= &  \Ent{X^{t+1}\ldots X^{t+k}R^{t}} - \Ent{X^{t+k}\ldots X^{t+1}R^{t+k}}.
\end{align}
The top term is \ref{lem:part:RetroNonUni}, the bottom is \ref{lem:part:blockH}
 and hence these expressions are equal.

\ref{lem:part:MemRetro}.
Showing this last form is equivalent is slightly more involved.
It may be proven by adapting and generalizing Theorem 1 of Mahoney et al.~\cite{MahoneyEC09} beyond causal states into general (possibly non-unifilar) memory $R$.

For some $l>k$, consider the expansions of the two terms:
\begin{align}
\Ent{R^t X^{t+1} \ldots X^{t+l} } & = \Ent{X^{t+1} \ldots X^{t+l} }  \nonumber \\
& \quad + \cEnt{R^t}{X^{t+1} \ldots X^{t+k} }, \\
\Ent{R^{t+k} X^{t+1} \ldots X^{t+l} } & = \Ent{X^{t+1} \ldots X^{t+l} } \nonumber \\
& \quad + \cEnt{R^{t+k}}{X^{t+1} \ldots X^{t+l} }.
\end{align}
The difference between these two terms is:
\begin{align}
\Delta & := \Ent{R^t X^{t+1} \ldots X^{t+l} } - \Ent{R^{t+k} X^{t+1} \ldots X^{t+l} }  \nonumber \\
& = \Ent{R^t X^{t+1} \ldots X^{t+k}} \nonumber \\
& \qquad + \cEnt{X^{t+k+1}\ldots X^{t+l}}{R^t X^{t+1}\ldots X^{t+k}} \nonumber \\
& \quad - \Ent{R^{t+k} X^{t+1}\ldots X^{t+k}} \nonumber \\
& \qquad - \cEnt{X^{t+k+1}\ldots X^{t+l}}{R^{t+k} X^{t+1}\ldots X^{t+k}}. \nonumber \\
\end{align}

Both the conditional entropy terms are the conditional entropy of the future of the pattern (steps $t\!+k\!+1$ to $t\!+\!l$) with respect to the part of the pattern (and memory) that they are conditioned on (steps $t$ to $t\!+\!k$).
This allows us to use the fact that $R$ is prescient to argue that $\cEnt{\future{X}^{t+k}}{R^{t+k} X^{t+1}\ldots X^{t+k}} = \cEnt{\future{X}^{t+k}}{R^{t+k}}$, since $R^{t+k}$ already contains all the information to predict $\future{X}^{t+k}$, the additional $X^{t+1}\ldots X^{t+k}$ are redundant --
a property of prescient memory known as {\em causal shielding}~\cite{Pearl00}.
Similarly, $\cEnt{\future{X}^{t+k}}{R^t X^{t+1}\ldots X^{t+k}} = \cEnt{\future{X}^{t+k}}{R^{t+k}}$  (see discussion within theorem~\ref{thm:generate}).
Thus, these two terms are the same, and we can hence simplify the expression to:
\begin{align}
\Delta & = \cEnt{R^t}{X^{t+1}\ldots X^{t+l}} - \cEnt{R^{t+k}}{X^{t+1}\ldots X^{t+l}}  \nonumber \\
& = \Ent{R^t X^{t+1} \ldots X^{t+k}} \nonumber \\
& \quad - \Ent{R^{t+k} X^{t+1}\ldots X^{t+k}}.
\end{align}
Thus, $\Delta = \chi_R(k)$ in the form given by \ref{lem:part:blockH}.
Since this equality is true for all $l>k$, we can then take the limit $l\to\infty$ (such that $X^{t+1}\ldots X^{t+l} \to \future{X}$) and arrive at
\begin{equation}
\chi_R(k) = \cEnt{R^t}{\future{X}} - \cEnt{R^{t+k}}{\future{X}},
\end{equation}
proving \ref{lem:part:blockH} and \ref{lem:part:MemRetro} are equivalent.
\end{proof}
\end{lemma}

\noindent
\inlineheading{Memory $k$-step crypticity.}
The crypticity of a pattern~\cite{MahoneyEC09,MahoneyEJC11} is given $\chi = \cEnt{S^t}{\future{X}}$.
This property can be generalized to a {\em memory crypticity} $\chi_R$, defined:
\begin{equation}
\chi_R = \cEnt{R^t}{\future{X}} - \cEnt{R^t}{\past{X}}.
\end{equation}
The second term subtracts any uncertainty in the memory state having observered the entire sequence of pattern to date.
For causal states $R=S$, $\chi_S = \chi$, since $\cEnt{S^t}{\past{X}} = 0$.

Using the form of $\chi_R(k)$ in lemma~\cref{thm:XrForm}(\ref{lem:part:PastFutMemH}), we see $\lim_{k\to\infty} \chi_R(k) = \chi_R$.
This motivates the naming of the quantity $\chi_R(k)$ as the {\em memory $k$-step crypticity}.
Moreover, in any of the forms listed in \ref{lem:part:PastFutMemH} when $R=S$, $\chi_S(k) =: \chi(k)$, the pattern's intrinsic $k-$step crypticity (defined in~\cite{MahoneyEC09} as the ``$k$-cryptic approximation.'').

We also remark that by treating the $k=0$ case an invitation to completely omit the terms ``$X^{t+1}\ldots X^{t+k}$'' in lemma~\cref{thm:XrForm}, then $\chi_R(0)=0$ in every form. (It is established similarly in~\cite{MahoneyEC09} that $\chi(0)=0$).
Physically (i.e.\ when taken with theorem~\cref{thm:generate}), this is a statement that any ``generator'' that produces nothing and does not change its memory is theoretically allowed to dissipate no work, regardless of what memory it has.

Despite not necessarily corresponding to causal states, nor necessarily updating in a unifilar manner,
 the memory $k-$step crypticity $\chi_R(k)$ shares a few useful properties with the pattern's intrinisic $k-$step crypticity $\chi(k)$, which we now prove:-

\begin{lemma}[$\chi_R(k)$ is non-negative and non-decreasing]
\label{lem:NonDecreasingCrypt}
For any choice of prescient memory $R$,
 $\chi_R(k)\geq 0$ for all $k\geq0$, and $\chi_R(k) \geq \chi_R(k')$ for all $k\geq k'$.
\begin{proof}
First, recall from theorem \cref{thm:XrForm}(\ref{cor:InfiniteFuture}) that
\begin{align}
\chi_R(k) & = \cEnt{R^t}{\future{X}^t} - \cEnt{R^{t+k}}{\future{X}^t}.
\end{align}

The entropy $\cEnt{R^{t+k}}{\future{X^t}} \geq \cEnt{R^{t+k+1}}{\future{X^t}}$.
This is most obviously seen by using stationarity to re-write the two terms as:
\begin{align}
\cEnt{R^{t+k}}{\future{X^t}} & =\; \cEnt{R^t}{X^{t-k+1}\ldots X^{t} \future{X^t}} \\
\cEnt{R^{t+k+1}}{\future{X^t}} & = \; \cEnt{R^t}{X^{t-k} X^{t-k+1} \ldots X^{t} \future{X^t}}.
\end{align}
Since the latter term is conditioned on the same variables as the former, plus an additional variable $X^{t-k}$, it can not be higher.
This non-increasing property then implies by induction $\cEnt{R^{t+k}}{\future{X}} \leq \cEnt{R^{t}}{\future{X}}$.
And hence we see that
\begin{align}
\chi_R(k) = \cEnt{R^t}{\future{X}} - \cEnt{R^{t+k}}{\future{X}} \geq 0
\end{align}
for all $k$.

Since $\chi_R(k) = \cEnt{R^t}{\future{X}^t} - \cEnt{R^{t+k}}{\future{X}^t}$ is the difference between a constant term and a non-increasing term, it  follows that $\chi_R(k)$ is non-decreasing.

\end{proof}
\end{lemma}

In the context laid out in our article, the above lemma may be interpreted physically:-
\begin{theorem}
\label{thm:NonDecreasePhysical}
For any generator with prescient memory $R$, the dissipative work cost of generation $W^k_{\rm diss}(R)$ is {\em always} non-negative.
In the special case where the process is implemented optimally (i.e.\ at the limit from Landauer's principle), a generator that produces $k$ steps of the pattern never dissipates less work than a generator that produces $k' < k$ steps of the pattern.
\begin{proof}
Recall that
\begin{equation}
\beta W^k_{\rm diss}(R) \geq~ \chi_R(k).
\end{equation}
From lemma~\ref{lem:NonDecreasingCrypt}, $\chi_R(k)\geq0$ and hence
\begin{equation}
\beta W^k_{\rm diss}(R) \geq~ 0.
\end{equation}

When implemented optimally (i.e.,\ at the limit from Landauer's principle), this inequality $\beta W^k_{\rm diss}(R) \geq~ \chi_R(k)$ is saturated.
In this regime, the monotonic non-decreasing nature of $\chi_R(k)$ (lemma~\ref{lem:NonDecreasingCrypt}) guarantees monotonic non-decreasing dissipation.
\end{proof}
\end{theorem}

We can also show that $\chi_R(k)$ is convex upwards for $k\geq 0$. To do this, we first prove the following lemma:

\begin{lemma}[Monotonically decreasing retrodiction]
\label{eq:MonoRetro}
If prescient memory $R$ can be used by a generator to produce a pattern,
 the quantity
\begin{equation}
\rho(k) := \begin{cases}
0 & \quad k = 0, \\
\cEnt{X^t}{R^t} & \quad k = 1, \\
\cEnt{X^t}{ X^{t+1} \ldots X^{t+k-1} R^{t+k-1} } & \quad k>1
\end{cases}
\end{equation}
is non-increasing with respect to $k$.
\begin{proof}
For causal states, an elegant proof of this presented as theorem 2 of \cite{MahoneyEJC11}.
However, as generic memory $R$ is not necessarily unifilar, their method can not be generalized here.
Instead, we supply an alternative proof that relies on the data-processing inequality, that states:
\begin{equation}
\Info{A}{B} \geq \Info{A}{f(B)}
\end{equation}
for random variables $A$ and $B$, and a map $f$ that acts on them.
(Intuitively, post-processing local data can not increase its non-local correlations).

If $R$ is memory that is physically used by a generator to produce a pattern one step at a time, there must be some function $f:X_{\rm dflt}R^t \mapsto R^{t+1} X^{t+1}$ corresponding to the update.
Thus, from the data-processing inequality
\begin{align}
\Info{X^t X_{\rm dflt}}{R^t} & \geq \Info{X^t X_{\rm dflt}}{X^{t+1}R^{t+1}}
\end{align}
Since $X_{\rm dflt}$ is uncorrelated with everything, we can omit it from both sides of the equation
\begin{equation}
\Info{X^t}{R^t} \geq \Info{X^t}{X^{t+1}R^{t+1}},
\end{equation}
which in terms of conditional entropy is
\begin{equation}
\cEnt{X^t}{R^t} \leq \cEnt{X^t}{X^{t+1}R^{t+1}}.
\end{equation}
Hence $\rho(2)\geq\rho(1)$.

Closely related to the update function $f$ is the function $f': X^{t+1}\ldots X^{t+k-1} X_{\rm dflt} R^{t+k-1} \mapsto X^{t+1}\ldots X^{t+k} R^{t+k}$, which describes the update in the presense of a stretch of the tape $X^{t+1}\ldots X^{t+k}$, which remains undisturbed and moreover does not affect the choice of update.
This is then a tensor product of the identity function on $X^{t+1}\ldots X^{t+k-1}$ with function $f$ on $X_{\rm dflt}R^{t+k-1}$; and so if $f$ is a valid positive map, so too must be $f'$.
Thus, we may once more use the data-processing inequality:
\begin{align}
\Info{X^t}{X^{t+1}\ldots X^{t+k-1} X_{\rm dflt} R^{t+k-1}} \hspace{-10em} & \nonumber \\
& \geq \Info{X^t}{X^{t+1}\ldots X^{t+k-1} X^{t+k} R^{t+k}}.
\end{align}
Taking into account that $X_{\rm dflt}$ has no correlations, and writing in terms of conditional entropies
\begin{align}
\cEnt{X^t}{X^{t+1}\ldots X^{t+k-1} R^{t+k-1}} \hspace{-10em} & \nonumber \\
& \leq \cEnt{X^t}{X^{t+1} X^{t+k} R^{t+k}}.
\end{align}
By induction, this proves the claim for $k>1$
Trivially, $0 \leq \cEnt{X^t}{R^t}$, because the term is an entropy.
Together, these statements show that $\rho$ monotonically increases for all $k\geq0$.
\end{proof}
\end{lemma}

\begin{lemma}[Memory-crypticity is convex upwards]
\label{lem:ConvexUpCrypt}
$\chi_R\left(k\right)$ is convex upwards with respect to $k$.
That is,
\begin{equation}
\chi_R\left(k\right) - \chi_R\left(k-1\right) \geq \chi_R\left(k'\right) - \chi_R\left(k'-1\right)
\end{equation}
when $k \leq k'$.
\begin{proof}
We can expand $\chi_R(k)$ in the form of lemma~\cref{thm:XrForm}(\ref{lem:part:blockH}):
\begin{align}
\chi_R\left(k\right) & = \Ent{X^{t+1}\ldots X^{t+k}R^{t}} \nonumber \\
& \quad - \Ent{X^{t+k}\ldots X^{t+1}R^{t+k}} \nonumber \\
& = \Ent{X^{t+k-1}\ldots X^{t+1}R^{t}} \nonumber \\
& \qquad + \cEnt{X^{t+k}}{X^{t+k-1}\ldots X^{t+1}R^{t}} \nonumber \\
& \quad - \Ent{X^{t+2}\ldots X^{t+k}R^{t+k}}  \nonumber \\
& \qquad - \cEnt{X^{t+1}}{X^{t+2}\ldots X^{t+k}R^{t+k}}.
\end{align}
Using stationarity and re-arranging:
\begin{align}
\chi_R\left(k\right) & = \Ent{X^{t+k-1}\ldots X^{t+1}R^{t}} \nonumber \\
& \qquad - \Ent{X^{t+1}\ldots X^{t+k-1}R^{t+k-1}}  \nonumber \\
& \quad + \cEnt{X^{t+k}}{X^{t+k-1}\ldots X^{t+1}R^{t}} \nonumber \\
& \qquad - \cEnt{X^{t}}{X^{t+1}\ldots X^{t+k-1}R^{t+k-1}}  \nonumber \\
& = \chi_R\left(k-1\right) + \cEnt{X^{t+k}}{X^{t+k-1}\ldots X^{t+1}R^{t}} \nonumber \\
& \quad - \cEnt{X^{t}}{X^{t+1}\ldots X^{t+k-1}R^{t+k-1}}.
\end{align}

Using the prescience of $R$, the second term
$\cEnt{X^{t+k}}{X^{t+k-1}\ldots X^{t+1}R^{t}} = \cEnt{X^{t+k}}{R^{t+k-1}}$ and, from stationarity, is equal to $\cEnt{X^{t+1}}{R^t}$.
Thus,
\begin{align}
\chi_R\left(k\right) & = \chi_R\left(k-1\right)  + \cEnt{X^{t+1}}{R^{t}} \nonumber \\
& \quad - \cEnt{X^{t}}{X^{t+1}\ldots X^{t+k-1}R^{t+k-1}}.
\end{align}

Thus we find by induction
\begin{align}
\chi_R\left(k\right) & = k \cEnt{X^{t+1}}{R^{t}} - \Gamma(k), \label{eq:CryptGamma}
\end{align}
where
\begin{align}
\Gamma(k) & = \cEnt{X^t}{R^t} + \sum_{i=1}^{k-1} \cEnt{X^{t}}{X^{t+1}\ldots X^{t+i}R^{t+i}} \nonumber \\
& = \sum_{i=0}^{k} \rho (i),
\end{align}
where $\rho(i)$ is the expression from lemma~\cref{eq:MonoRetro}, which has been established to be positive and monotonically non-decreasing.
The sum $\Gamma(k)$ of such monotonotically non-decreasing terms must hence be convex downward for $k\geq0$.
Since $\chi_R\left(k\right)$ is then the difference between a linear contribution and a convex downward contribution, it must be convex upwards.

This property remains true when we extend the domain of $\chi_R(k)$ to include $\chi_R(0)=0$ ($\Gamma(0)\!=\!\rho(0)\!=\!0$).
\end{proof}
\end{lemma}

When taken together with theorem~\ref{thm:generate}, the above property has the following physical implication:-
\begin{theorem}
\label{lem:SublinearCrypt}
\label{thm:ThermoCausality}
For any generator optimally implementing prescient memory $R$,
 the minimum work investment {\em per step} of the pattern produced
 is never larger if the generator produces a larger string of the pattern in any given run.
That is, for a given choice of prescient memory $R$, it is thermodynamically better to produce as much of the pattern as possible at once.
\begin{proof}
Since $\chi_R\left(k\right)$ is convex upwards (lemma~\cref{lem:ConvexUpCrypt}), including its extension to $\chi_R\left(0\right)=0$,
 this immediately implies
\begin{align}
\label{eq:SubLinear}
\dfrac{\chi_R\left(k\right)}{k} \geq \dfrac{\chi_R\left(k'\right)}{k'} ~\mathrm{when}~ 1\leq k \leq k'.
\end{align}

Taken together with theorem~\cref{thm:generate}, this immediately implies the claim.

\end{proof}
\end{theorem}

We can now show that for any choice of $R$, either $\chi_R(k)=0$ for all $k\geq1$, or $\chi_R(k)\neq0$ for all $k\geq1$.
That is, it is not possible for only some of $\chi_R(k\geq1)$ to be zero.

\begin{lemma}[All zero]
\label{lem:AllZero}
For a given choice of $R$, if $\chi_R(k)=0$ for any $k\geq1$, then $\chi_R(k) = 0$ for all $k\geq1$.
Also, if $\chi_R = \lim_{k\to\infty} \chi_R(k)= 0$, then $\chi_R(k) = 0$ for all $k\geq1$.
Moreover, if $\chi_R(k)=0$ for any $k\geq1$ then $\chi_R = \lim_{k\to\infty} \chi_R(k) \to 0$.
\begin{proof}
First consider $k' < k$. From lemma~\cref{lem:NonDecreasingCrypt}, $\chi_R$ is non-decreasing and positive.
Hence $\chi_R(k)=0$ implies that $\chi_R(k' ) = 0$ for all $k'\leq k$, .

Now consider for $k' > k$.
From theorem~\cref{lem:SublinearCrypt}
\begin{equation}
\dfrac{\chi_R\left(k\right)}{k} \geq \dfrac{\chi_R\left(k'\right)}{k'} \mathrm{~when~} k \leq k'.
\end{equation}
and since $\chi_R(k)=0$,
\begin{equation}
0 \geq \dfrac{\chi_R\left(k'\right)}{k'}
\end{equation}
implying that $\chi_R\left(k'\right) \leq 0$.
But lemma~\cref{lem:NonDecreasingCrypt} states $\chi_R\left(k'\right)\geq0$, and hence we may conclude that $\chi_R\left(k'\right)=0$.
Thus, if $\chi_R(k)=0$ for one value of $k\geq1$, $\chi_R(k)=0$ for all values of $k\geq1$.

The second sentence of the claim is simpler to prove.
Again, lemma~\cref{lem:NonDecreasingCrypt} states that $\chi_R$ is non-decreasing and positive.
If its limit $\chi_R = \lim_{k\to\infty} \chi_R(k)= 0$, then $\chi_R(k)=0$ for all $k$, since this limit is approached from below.

Finally, to show the last sentence of the claim, we more carefully pick at ineq.~\eqref{eq:SubLinear} of theorem~\cref{lem:SublinearCrypt}.
Writing as
\begin{align}
k \dfrac{\chi_R\left(k'\right)}{k'} \geq \chi_R\left(k\right), \nonumber \\
k \times 0 \geq \chi_R\left(k\right),
\end{align}
where $\dfrac{\chi_R\left(k'\right)}{k'}$ is strictly $0$ for all $k$  (not, say, some finite expression that becomes vanishingly small with $k$).
Only then may we safely take the limit $k\to\infty$, such that $\lim_{k\to\infty} (0 \times k) = 0$ and hence $\lim_{k\to\infty} \chi_R\left(k\right) \leq 0$.
It thus follows that $\chi_R =0$.

\end{proof}
\end{lemma}

\begin{corollary}[All non-zero] 
\label{lem:NoneZero}
If for some $k\geq 1$, $\chi_R(k)>0$, then $\chi_R(j)>0$ for all $j\geq1$, and $\chi_R= \lim_{k\to\infty} \chi_R(k) > 0$.
Likewise, if $\chi_R = \lim_{k\to\infty} \chi_R(k) > 0$, then $\chi_R(j)>0$ for all $j\geq1$.
\begin{proof}
This follows by contradiction with \mbox{lemma~\ref{lem:AllZero}.}
Suppose $\chi_R(k) > 0$ but either for some $k'\geq 1$, $\chi_R(k')=0$ or $\chi_R = 0$.
Lemma~\cref{lem:AllZero} states that this implies $\chi_R(k)=0$, immediately leading to contradiction.
Likewise, if $\chi_R > 0$, but that there was some finite $k\geq1$ such that $\chi_R(k) = 0$,
 from lemma~\cref{lem:AllZero} $\chi_R=0$, leading to contradiction.
\end{proof}
\end{corollary}

This allows us make our first main result:-
\begin{result}[Excessive information causes dissipation]
\label{res:CrypticDissipation}
$W^k_{\rm diss} > 0$ whenever $\Info{R^t}{\past{X}} >  \Info{\past{X}}{\future{X}}$.
\begin{proof}
Using the form of $\chi_R$
 from lemma~\cref{thm:XrForm}(\ref{lem:part:RetPredI})
\begin{align}
\label{eq:appNostalgia}
\beta W^k_{\rm diss} \geq~ & \chi_R(k) = \Info{X^{t-k+1}\ldots X^{t}}{R^{t}} \nonumber \\
& \quad - \Info{X^{t+k}\ldots X^{t+1}}{R^{t}}.
\end{align}

In the limit of $k\to\infty$
\begin{align}
\lim_{k\to\infty} \chi_R(k) = \Info{\past{X}}{R^t} - \Info{\future{X}}{R^t},
\end{align}
and since $R$ is prescient, we can replace the last term with $\Info{\future{X}}{\past{X}}$:
\begin{align}
\lim_{k\to\infty} \chi_R(k) = \Info{\past{X}}{R^t} - \Info{\future{X}}{\past{X}}.
\end{align}

Suppose $\Info{\past{X}}{R^t} \neq \Info{\future{X}}{\past{X}}$ such that $\chi_R \neq 0$.
Then by corollary~\ref{lem:NoneZero}, $\chi_R(k) > 0$ for {\em all} $k$.
Putting this into theorem~\ref{thm:generate}, it then immediately follows that $\beta W^k_{\rm diss} > 0$ for all $k$.
\end{proof}
\end{result}

\begin{result}[Simpler is thermodynamically better]
\label{res:Simpler}
For generating $k$ steps of any given pattern $\pastfuture{X}$,
 the generator's dissipative work cost is minimized by
 choosing prescient memory $R$ to be in one-to-one correspondence with the pattern's causal states $S$.
\begin{proof}
Consider then theorem~\cref{thm:generate}, namely $\beta W^k_{\rm diss}(R) \geq~ \chi_R(k)$.
If implemented at the theoretical optimal limit dictated by Landauer's principle, then equality holds.
It then follows that to achieve the optimal thermodynamic performance, one should minimize $\chi_R(k)$.

From lemma~\cref{thm:XrForm}(\ref{lem:part:PredRetH}),
\begin{align}
\chi_R(k) =
& \cEnt{X^{t+1}\ldots X^{t+k}}{R^t} \nonumber \\
& \quad - \cEnt{X^{t+k}\ldots X^{t+1}}{R^{t+k}}.
\label{eq:dissWorkSIB}
\end{align}
The definition of prescience for $R^t$ tells us that
 $\cPr{\future{X}}{R^t} = \cPr{\future{X}}{\past{X}}$,
 and hence for any string of length $k$, $\cEnt{X^{t+1}\ldots X^{t+k}}{R^t} = \cEnt{X^{t+1}\ldots X^{t+k}}{\past{X}}$.
For all choices of memory $R$ (including when it is in one-to-one correspondence with the causal states $S$) this quantity is the same,
 and the minimization can be performed entirely by maximizing the second term of equation~\eqref{eq:dissWorkSIB}.

Prescience implies an important condition on memory $R$:
 namely that it encodes a {\em refinement} of causal states:
 no two past histories $\past{x}$ and $\past{x'}$ can be mapped to the same state $r$ if they belong to two separate causal states.
This property immediately follows from our definition of prescience:
 if $\cPr{\future{X}}{\past{X}=\past{x}}\neq\cPr{\future{X}}{\past{X}=\past{x}'}$, it is clearly impossible for $\cPr{\future{X}}{R=r}$ to be equal to both.
(See also Lemma 7 of \cite{ShaliziC01}.)
This refinement property implies the existence of a deterministic map $\Phi: \{r\} \to \{s\}$, such that $\Phi(R) = S$.
Then, we can apply the data processing inequality $\Info{X^{t+1}\ldots X^{t+k}}{R^{t+k}} \leq \Info{X^{t+1}\ldots X^{t+k}}{\Phi\!\left(R^{t+k}\right)}
$.
Hence,
\begin{equation}
\Info{X^{t+1}\ldots X^{t+k}}{R^{t+k}} \leq \Info{X^{t+1}\ldots X^{t+k}}{S^{t+k}}.
\end{equation}
Expanding these mutual informations gives:
\begin{align}
&\Ent{X^{t+1}\ldots X^{t+k}} - \cEnt{X^{t+1}\ldots X^{t+k}}{R^{t+k}}  \nonumber \\
& \qquad \leq \Ent{X^{t+1}\ldots X^{t+k}} - \cEnt{X^{t+1}\ldots X^{t+k}}{S^{t+k}},  \nonumber \\
&\cEnt{X^{t+1}\ldots X^{t+k}}{S^{t+k}} \geq \cEnt{X^{t+1}\ldots X^{t+k}}{R^{t+k}}.
\end{align}

Thus, we see that using memory in one-to-one correspondence with the causal states $S$ minimizes eq.~\eqref{eq:dissWorkSIB}.
That is for any pattern,
\begin{equation}
\chi_R(k) \geq \chi_S(k) \mathrm{~for~all~} R, k.
\end{equation}

This proves the claim of the result.
\end{proof}
\end{result}

We remark that this lower bound is not trivially saturated for all $R$.
This is proven by the example in the article; where choosing $R$ to not correspond to causal states in general resulted in increased dissipation.

\begin{result}
\label{res:BarbedArrow}
 A thermodynamically reversible cycle of pattern generation and extraction is impossible for any pattern where $C_\mu > \Info{\past{X}}{\future{X}}$,
 where $C_\mu$ is the pattern's {\em statistical complexity}.
\begin{proof}
This is a corollary of \cref{res:CrypticDissipation,res:Simpler}.
Consider using memory in one-to-one correspondence with the causal states.
In the limiting case of $k\to\infty$:
\begin{equation}
\lim_{k\to\infty} \beta W^k_{\rm diss} \geq \chi_S = \Info{\past{X}}{S} - \Info{\past{X}}{\future{X}}.
\end{equation}
Because causal states can be synchronized, $\Info{\past{X}}{S} = \Ent{S} = C_\mu$, and hence
\begin{equation}
\chi_S = C_\mu - \Info{\past{X}}{\future{X}}.
\end{equation}
When $C_\mu > \Info{\past{X}}{\future{X}}$, $\chi_S > 0$, and it follows from \cref{res:CrypticDissipation} that for any $k$ there will be some dissipation with this choice of memory.
\Cref{res:Simpler} then tells us that no other choice of memory can do better than this.
It hence follows that when $C_\mu > \Info{\past{X}}{\future{X}}$, any attempt to generate a pattern will result in some work dissipation.
\end{proof}
\end{result}

\begin{example}[Further details of worked example]
\label{EG:PertCoinProof}
For a ``last $N$'' generator that produces $k$ steps of the {\em perturbed coin} pattern,
 using as its internal memory configuratoins in one-to-one correspondence with the last $N$ outputs of the sequence,
 the amount of dissipation $W^k_{\rm diss}$ is bounded from below by:
\begin{equation}
\beta W^k_{\rm diss} \geq
\begin{cases}
k h\!\left(p\right)  & \mathrm{when~} k \leq N \\
N h\!\left(p\right)  & \mathrm{when~}k > N,
\end{cases}
\end{equation}
where $h\!\left(p\right)$ is the {\em binary entropy} defined as
\begin{equation}
h\!\left(p\right) := -p \log p - \left(1\!-\!p\right) \log \left(1\!-\!p\right).
\end{equation}
\begin{proof}
From theorem~\cref{thm:generate}, $\beta W^k_{\rm diss} \geq \chi_R(k)$.
We now derive the form of $\chi_R(k)$ for the last-$N$ machine.

We begin with the form in lemma~\cref{thm:XrForm}(\ref{lem:part:PredRetH}):
\begin{align}
\chi_R(k) & = \cEnt{X^{t+1}\ldots X^{t+k}}{R^t} \nonumber \\
& \quad - \cEnt{X^{t+k}\ldots X^{t+1}}{R^{t+k}}.
\label{eq:MarkovCoinXr}
\end{align}
Let us evaluate both terms.

First,
\begin{align}
\cEnt{X^{t+1}\ldots X^{t+k}}{R^t} \hspace{-5em}  & \nonumber\\
 & = \cEnt{X^{t+1}}{R^t} + \cEnt{X^{t+2}}{R^t X^{t+1}} \nonumber \\
 & \hspace{1em} + \ldots
 + \cEnt{X^{t+k}}{R^t X^{t+1} \ldots X^{t+k-1}}.
\end{align}
Since $R$ is prescient, and all the machines of the ``last $N$'' type are unifilar (such that $\cEnt{R^{t+j}}{R^{t}X^{t+1}\ldots X^{t+j}} = 0$ for any integer $j$),
it follows (using stationarity in the second step) that:
\begin{align}
\cEnt{X^{t+1}\ldots X^{t+k}}{R^t} \hspace{-5em}  & \nonumber\\
 & = \cEnt{X^{t+1}}{R^t} + \cEnt{X^{t+2}}{R^{t+1}} \nonumber \\
 & \hspace{1em} + \ldots
 + \cEnt{X^{t+k}}{R^{t+k-1}}, \nonumber \\
 & = k\cEnt{X^{t+1}}{R^t}.
\end{align}
Moreover, since $R$ is prescient,
\begin{equation}
\label{eq:NostRHSPerturbed}
\cEnt{X^{t+1}\ldots X^{t+k}}{R^t} = \cEnt{X^{t+1}\ldots X^{t+k}}{S^t} = k h\!\left(p\right).
\end{equation}

Now we evaluate the second term $\cEnt{X^{t+1}\ldots X^{t+k}}{R^{t+k}}$.
When $k\leq N$,
 since $R$ encodes all the available information about the last $N$ outputs,
 $\cEnt{X^{t+1}\ldots X^{t+k}}{R^{t+k}} = 0$.
Thus,
\begin{equation}
\label{eq:PCChiSmallN}
\chi_R(k) = k h\!\left(p\right) \mathrm{~when~} k \leq N.
\end{equation}

Next, consider the case where $k>N$.
We expand
\begin{align}
\cEnt{X^{t+1}\ldots X^{t+k}}{R^{t+k}} \hspace{-8em} & \nonumber\\
& = \cEnt{X^{t+k-N}\ldots X^{t+k}}{R^{t+k}} \nonumber \\
& \hspace{1em} + \cEnt{X^{t+1}\ldots X^{t+k-N-1}}{X^{t+k-N}\ldots X^{t+k} R^{t+k}} \nonumber \\
& = \cEnt{X^{t+1}\ldots X^{t+k-N-1}}{X^{t+k-N}\ldots X^{t+k} R^{t+k}},
\end{align}
where the first term was eliminated because $R^{t+k}$ contains all the information about the preceeding $N$ outputs of the pattern.

For short-hand, we write $j = k - N + 1$.
We can show that knowledge of $R^{t+k}$ gives no further information beyond that in $X^{t+j}\ldots X^{t+k}$ for the purpose of determining the values of $X^{t+1}\ldots X^{t+j-1}$.
This is seen by performing the following expansion:
\begin{align}
\cEnt{X^{t+1}\ldots X^{t+j-1}R^{t+k}}{X^{t+j}\ldots X^{t+k}} \hspace{-15em} & \nonumber\\
& = \cEnt{R^{t+k}}{X^{t+j}\ldots X^{t+k}} \nonumber \\
& \quad + \cEnt{X^{t+1}\ldots X^{t+j-1}}{X^{t+j}\ldots X^{t+k} R^{t+k}}\nonumber \\
& = \cEnt{X^{t+1}\ldots X^{t+j-1}}{X^{t+j}\ldots X^{t+k}}\nonumber \\
& \quad + \cEnt{R^{t+k}}{X^{t+1}\ldots X^{t+k}}.
\end{align}
Since $k>N$, both
 $\cEnt{R^{t+k}}{X^{t+j}\ldots X^{t+k}} = 0$
 and $\cEnt{R^{t+k}}{X^{t+1}\ldots X^{t+k}} = 0$,
 since both strings have (at least) the $N$ required outputs to fix the value of $R^{t+k}$.
Thus,
$\cEnt{X^{t+1}\ldots X^{t+j-1}}{X^{t+j}\ldots X^{t+k} R^{t+k}}
 = \cEnt{X^{t+1}\ldots X^{t+j-1}}{X^{t+j}\ldots X^{t+k}}$.

We can now use the Markovian nature of the perturbed coin pattern to further simply the above expression.
For any Markovian process, by definition $\cEnt{X^{t+l}}{X^{t+1}\ldots X^{t+l-1}} = \cEnt{X^{t+l}}{X^{t+l-1}}$ for any $l\in\mathbb{Z}^+$.
From this condition, it is possible to obtain the inverse statement for any $1 < n < m$:
\begin{align}
\cEnt{X^{t+1}\ldots X^{t+n-1}}{X^{t+n}\ldots X^{t+m}} \hspace{-10em} & \nonumber \\
& = \cEnt{X^{t+1}\ldots X^{t+n-1}}{X^{t+n}}.
\end{align}

We use this to further simplify
\begin{align}
\cEnt{X^{t+1}\ldots X^{t+j-1}}{X^{t+j}\ldots X^{t+k}}  \hspace{-15em} & \nonumber\\
& = \cEnt{X^{t+j-1}}{X^{t+j}\ldots X^{t+k}} + \ldots \nonumber \\
&\quad + \cEnt{X^{t+1}}{X^{t+2}\ldots X^{t+k}} \nonumber \\
& = \cEnt{X^{t+j-1}}{X^{t+j}} + \ldots \nonumber \\
&\quad + \cEnt{X^{t+1}}{X^{t+2}} \nonumber \\
& = (j - 1)\cEnt{X^t}{X^{t+1}},
\end{align}
where the final step follows from the stationary nature of the process.
Noting that $j-1 = k - N$,
 and directly calculating from the description of the perturbed coin process that $\cEnt{X^t}{X^{t+1}}=h\!\left(p\right)$,
 we thus calculate that when $k>N$, the second term of equation~\eqref{eq:MarkovCoinXr} is
\begin{equation}
\cEnt{X^{t+1}\ldots X^{t+k}}{R^{t+k}} = \left(k - N\right) h\!\left(p\right).
\end{equation}
Taking the difference of eq.~\eqref{eq:NostRHSPerturbed} and the above yields:
\begin{equation}
\label{eq:PCChiLargeN}
\chi_R(k) =N h\!\left(p\right) \mathrm{~when~} k > N.
\end{equation}

From theorem~\cref{thm:generate}, $\beta W^k_{\rm diss} \geq \chi_R(k)$. Thus, equations~\eqref{eq:PCChiSmallN} and ~\eqref{eq:PCChiLargeN} hence prove the claim.

\end{proof}
\end{example}

\begin{figure}[tbh]
\includegraphics[width=0.425\textwidth]{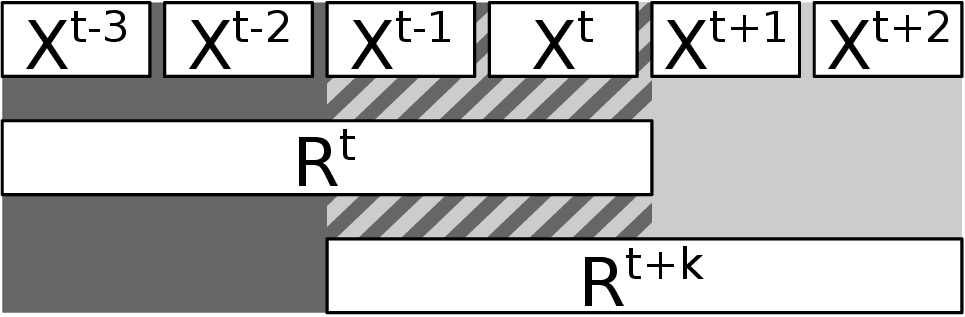}
\caption{
\label{fig:RevLastN}
\caphead{Logical reversibility in the last~$N$ machine.}
(Shown here for a last $N=4$ machine updating with stride $k=2$.)
The last $N$ machine's memory can be viewed as a window on the pattern of width $N$ that moves forwards by $k$ steps per cycle.
The darker region indicates information stored in $R^t$, and the lighter region the information stored in $R^{t+k}$.
Information corresponding to steps $X^{t-1}$ and $X^{t}$ (dashed region) exists in the machine's memory both before and after update.
The part of the update pertaining to this information can be performed in a logical reversible manner.
}
\end{figure}

That dissipation is limited by $k$ for $k<N$ and $N$ for $k\geq N$ may be understood in terms of the logical reversibility.
The memory $R$ of a last $N$ machine may be visualized (\cref{fig:RevLastN}) as a moving window of length $N$ scanning over a pattern, advancing by $k$ steps per update. 
When $k<N$, parts of this window overlap before and after the update: there is shared information contained within both $R^t$ and within $R^{t+k}$ pertaining to steps $X^{t-k+1}\ldots X^t$ that can be updated using a logically reversible operation.
On the other hand, when $k\geq N$, this overlap between $R^t$ and $R^{t+k}$ completely vanishes, and so the entire memory must be updated by logically irreversible processes.

One implementation of this would be to represent the last $N$ machine's memory by a compound structure $R^t=X^{t-N+1} \ldots X^{t-N+k} X^{t-N+k+1}\ldots X^{t}$ with $N$ {\em registers}, each storing information about one step of the pattern.
When updating by $k < N$, one can first cyclically permute the contents of the registers (equivalently, relabel their indices) such that $\tilde{R}^t = X^{t-N+k+1}\ldots X^{t} X^{t-N+1} \ldots X^{t-N+k}$.
This is an intrinsically reversible operation that requires no work investment.
As the desired final memory state is $R^{t+k} = X^{t-N+k+1}\ldots X^{t} X^{t+1} \ldots X^{t+k}$,
 the first $N-k$ registers of the memory in $\tilde{R}^t$ already have their correct values, 
 and work only needs to be invested to bring the final $k$ registers of the memory up to date.
On the other hand, when $k\geq N$, every register in the memory will require updating, and the cost will hence be bounded by the size of the memory $N$ rather than the size of the update $k$.

\newpage
\inlineheading{Physical example: the trajectory formalism.} 
For illustrative purposes, we present a physical model for the cycle of generation and extraction of the perturbed coin pattern.
We shall employ a subset of the {\em trajectory formalism}~(see e.g.~\cite{Alicki79,AlickiHHH04,Kieu06,QuanD08,KimSDU11,Aberg13,YungerHalpernGDV15,BrowneGDV14} among many), but stress that this framework is just one arbitrary choice from many thermodynamic models. 
The bounds derived in this article, being information-theoretic in origin, hold for any model of thermal interaction which defines heat and work in a way that is consistent with Landauer's principle.

We provide a few key details of this framework.
Consider a system with a finite number of well-defined energy levels (i.e.\ with a Hamiltonian $\mathcal{H} = \{E_1 \ldots E_N \}$) and a (classical) state, reflecting the occupation probabilities $\{P_1\ldots P_N\}$ that the system is in a particular energy level.
The energy of the system may change in one of two ways: 
 {\bf 1.}\ changes in the Hamiltonian, at fixed occupation probability; 
 {\bf 2.}\ changes in the occupation probability under a fixed Hamiltonian. 
We express this energy change differentially as
\begin{equation}
dU = \sum_{i=1}^N P_i dE_i + \sum_{i=1}^N E dP_i.
\label{eq:FirstLawClassical}
\end{equation}

The first type of energy change can be induced by some choice of time-varying external parameter (i.e.\ force), and we shall require that the change in force is independent of the system's state to ensures that there is no unaccounted-for feedback.

We will place more restrictive conditions on the allowed energy exchanges of the second type.
Namely, we only admit {\em thermalizing interactions} -- such that all transformations on the state necessarily take it closer to the Gibbs state for a given Hamiltonian 
(that is, the state where $P_i= \frac{e^{-\beta E_i}}{\sum_j e^{-\beta E_j}}$ for all $i$).
This ensures consistency with the second law of thermodynamics%
\footnote{
If additional behavioural constraints are imposed, the higher moments of the heat and work distributions' statistics can also be made to match physically-expected behaviour.
(E.g.\ imposing {\em detailed balance} ensures consistency with {\em fluctuation theorems}~\cite{Jarzynski97,Crooks99}).
A general description of models that allow this is in the trajectory formalism is in appendix A of \cite{YungerHalpernGDV15}.
}.

For the purpose of the example in this article, we shall only consider {\em quasistatic} protocols, in which whenever the heat bath is coupled to the system, the system is allowed to reach perfect thermal equilibrium, and so remains in the Gibbs state associated with the Hamiltonian.
This will trivially satisfy this requirement.
There is some indication~\cite{BrowneGDV14} that the cost of a quasistatic protocol bounds the actual finite-time cost reasonably tightly.

When all the above conditions are met, eq.~\eqref{eq:FirstLawClassical} becomes somewhat like the {\em first law}, and the trajectory formalism allows us to associate the first term $\sum_i P_i dE_i$ with {\em work} and the second term $\sum_i E_i dP_i$ with {\em heat}\;\footnote{
See e.g.\ \cite{WeimerHRSM08,VinjanampathyA16,GooldHRRS16} for discussion of the additional considerations that must be taken into account before one can also make this equivalence in the quantum regime. 
In this article, we do not need to assert a quantum definition of work.}.
To find the total work cost of a protocol, one typically integrates over a series of infinitesimal contributions.

For the purpose of calculating costs of a pattern-manipulating protocol within this framework, we now prove the following lemma:
\begin{lemma}[Work cost in a two-level system]
\label{lemma:TwoLevelLandauer}
In the trajectory formalism, for a two-level system that initially and finally is subject to a degenerate Hamiltonian, 
 there is a quasistatic procedure that transforms it from state $(q, 1-q)$ with binary entropy $h(q)$ to the state $(p, 1-p)$ with binary entropy $h(p)$, at a work cost given by
\begin{equation}
W = \kB T \left[ h(q) - h(p) \right],
\end{equation}
when the system has access to a thermal reservoir at temperature $T$.
\begin{proof}
We shall constructively provide a mechanism with this cost consisting of three stages.
Let $E_1$ and $E_2$ be the values of the first and second energy levels respectively, such that initially $E_1 = E_2 = 0$.

First, in thermal isolation, we change the Hamiltonian such that $E_2 = E_q$ where $E_q$ satisfies $q= \frac{1}{1+ e^{-\beta E_q}}$.
This produces a Hamiltonian where $(q, 1-q)$ is the associated Gibbs state.
Since the second level was initially populated with probability $(1-q)$, changing the Hamiltonian has a work cost (resp. gain if $q<\frac{1}{2}$) of $W_1 = (1-q)E_q$.

Next, we connect the system to a thermal bath. This has no effect on the occupation probabilities or Hamiltonian (and thus no associated work or heat cost).
We slowly change the second energy level to a value $E_2=E_p$ that satisfies $p = \frac{1}{1+ e^{-\beta E_p}}$.
When this is done quasistatically, and noting that $dE_1 = 0$ at all times, we find the total work exchange in this stage of the protocol is
\begin{align}
W_2 & =  \int_{E_q}^{E_p} \!P_2\;dE_2 = \int_{E_q}^{E_p} \! \dfrac{e^{-\beta E_2}}{1+e^{-\beta E_2}}\; dE_2 \nonumber\\
& = \frac{1}{\beta}\left( \ln p - \ln q \right)
\end{align}
where we made the substitution $u=1+e^{-\beta E_2}$ to solve the integral.

For the third and final stage of the protocol, we disconnect the system [now in state $(p, 1-p)$] from the thermal bath, and lower the second energy level back down to $E_2=0$. 
This induces a work exchange of $W_3 = -\left(1\!-\!p\right)E_p$.
Summing the contributions from the three parts of the protocol $W = W_1 + W_2 + W_3$, the total work exchange is:
\begin{equation}
W = (1\!-\!q)E_q - (1\!-\!p)E_p + \frac{1}{\beta}\left(\ln p -\ln q\right).
\end{equation}
However, since $E_q = \frac{1}{\beta}\left[\ln q - \ln\left(1-q\right)\right]$ (and likewise for $p$ and $E_p$), we may rewrite the work cost
\begin{align}
W & = \frac{1}{\beta}\left[p\ln p + \ln\left(1-p\right) - q \ln q - \left(1\!-\!q\right)\ln\left(1\!-\!q\right)\right], \nonumber\\
& = \kB T \left[ h(q) - h(p) \right], \label{eq:LandauerTwoLevel}
\end{align}
proving the claim.
\end{proof}
\end{lemma}
It follows from conservation of energy (or can be shown directly using a calculation similar to the lemma above) that the heat transferred into the heat bath during this transaction must be equal to the work invested, since the average internal energy has not changed between the initial and final states.

Unlike the other lemmata in this article, here we have not assumed that Landauer's principle holds -- the difference in entropies has appeared emergently from the protocol within the framework.
However, we also have not explicitly proved Landauer's principle, since this would require a minimization over all possible protocols. 
Rather, what is shown is the existence of a mechanism within the trajectory framework that performs the above transformation in a manner that saturates Landauer's bound.
If we consider the special case of $q = \frac{1}{2}$ and $p=1$, then the above lemma yields the famous bit-reset cost of $\kB T \ln 2$. Likewise by setting $q=1$ and $p=\frac{1}{2}$, we have the Szilard engine output $-\kB T \ln 2$.

\vspace{0.5em}
\inlinesubheading{Pattern manipulation within the trajectory formalism.}
With this in mind, we may now provide a model for the generator of the perturbed coin pattern.
Let us analyse the simplest possible example, where the pattern generator that writes one step at a time and has an internal memory configuration corresponding to the two causal states $s_H$ and $s_T$ (i.e.\ behaves according to \cref{fig:PerturbedCoin}).
We shall make an extra ``ancilla'' bit of memory available to the generator, but must take care that it has explicitly been reset by the end of the procedure.
As in \cref{fig:SimExtCycle}, we make baths at temperatures $T_G$ and $T_E$ available to the generator and extractor respectively, and assume that the energy investment required to make the work-like energy exchanges comes from some mutually-available work reservoir (battery).
This process may be viewed as an elaboration of the concepts mentioned in briefly in \cref{fig:UpdateTape,fig:UpdateState}, but for the specialized case of a particular pattern, running with particular memory, at a particular stride.

Let the system on the tape be a two-level system, initially configured according to a predefined default distribution $X_{\rm dflt} = (q, 1-q)$.
The first stage of the protocol will be to set the system from $X_{\rm dflt}$ to an intermediate state $(p, 1-p)$, where the value of $p$ is the ``swap probability'' determined by the perturbed coin process (i.e.\ has exactly the same meaning as in \cref{fig:PerturbedCoin}).
We immediately see that the work cost of this (interacting with a bath at inverse temperature $\beta_G$) is what we have just calculated in lemma~\cref{lemma:TwoLevelLandauer}: $\frac{1}{\beta_G} \left[\Ent{X_{\rm dflt}}- h(p)\right]$.
This cost corresponds to $W^1_{\rm gen}(S)$ of eq.~\ref{eq:GenerateCost}.

\enlargethispage{\baselineskip}
Now, let us consider the system and the memory together.
The memory has some state $S^t = s_h$ or $s_T$.
A controlled-not (CNOT) operation may be applied to the state of the system on the tape, such that if the memory was in state $S^t=s_h$, nothing happens [$X$ remains configured as $(p, 1-p)$], but if $S^t = s_T$ the occupation probabilities are flipped putting the tape system into the state $(1-p, p)$.
It is well-established that such purely reversible operations can be implemented at no net cost~\cite{Bennett82} (one can think of it much like a relabelling of energy levels).
After this operation, the system on the tape will have been encoded with statistics $X^{t+1}$ appropriate to the pattern.

Next, we must update the memory, to ensure that upcoming tape systems can also be set into the correct statistics (including appropriate correlations with $X^{t+1}$).
To do this, we take the ``ancilla'' bit [initially in pure state $(1,0)$], and apply another reversible CNOT operation on it, controlled by the state of the patterned tape $X^{t+1}$.
Next, we (reversibly) swap the state of the ancilla bit with our main memory bit.
Since these two procedures are both reversible, they do not contribute to any work or heat costs.

Let us summarize the random variables describing the state of all three systems at this point in time: the main memory is configured according to $S^{t+1}$, the ancilla to $S^{t}$ and the tape to $X^{t+1}$.
To finish the generation procedure in a manner that accounts for all potentially useful thermal resources, we must reset the ancilla from $S^t$ back to its initial pure state.
In the simple Markovian example of the perturbed coin pattern, $S^{t+1}=X^{t+1}$, and all the useful information the tape contains regarding how to reset the ancilla is already encoded in the main memory [entropically: $\cEnt{S^t}{S^{t+1}X^{t+1}} = \cEnt{S^t}{S^{t+1}}$].
This means we can at this point emit the patterned tape from the generating device, and consider the cost of resetting the ancilla from $S^t$ to $(1,0)$ only using knowledge in the main memory ($S^{t+1}$).

Clearly, except when $p=\frac{1}{2}$, there is some correlation between $S^t$ and $S^{t+1}$.
Thus, the first stage of our memory reset is to decorrelate the two systems, by applying another reversible CNOT gate on the ancilla, controlled by the state of the main memory.
This will set the ancilla into the state $(1-p, p)$ independent of the current value of the memory (this works because of the symmetry of the perturbed coin pattern; there was a probability $(1-p)$ that $S^{t+1}$ and $S^t$ are the same, and a probability $p$ that they were different). 

Now, we may use the protocol in lemma~\cref{lemma:TwoLevelLandauer} to take the ancilla from state $(1-p, p)$ back to the pure state $(1, 0)$ at a work cost of $\kB T_G h(p)$ (i.e.\ using a thermal bath at the same temperature $T_G$ as before).
This particular work cost is an example of $W^1_{\rm diss}(S)$ (see eq.~\eqref{eq:cEntExplicit}), as has been discussed extensively throughout the article and appendices.

Thus, the memory is now perfectly correlated with the patterned tape emitted, and ready to accept the next system on the tape and continue generating the pattern. 
The generation stage is hence complete, 
 requiring a total work cost of  $\frac{1}{\beta_G}\left(\Ent{X_{\rm dflt}} - h(p) + h(p)\right) = \kB T_G \Ent{X_{\rm dflt}}$.

Now consider the behaviour of the extractor, which we assume also has access to a heat bath at (possibly) different temperature $T_E$.
Initially, the system on the incoming tape is $X^{t+1}$ is, and the internal memory is in state $S^t$.
The first stage is to reversibly swap the state of the tape system with the memory state. 
Since the perturbed coin process is Markovian and $X^{t+1} = S^{t+1}$, this alone will ensure that the memory is in the correct state to anticipate future the extraction of upcoming parts of the pattern.
Now, we can apply a (reversible) CNOT on the system on the tape (currently in state $S^t$) controlled by the system in memory $S^{t+1}$, noting that as above there was a probability $1-p$ that the systems are the same, and of $p$ that they are different.
The system on the tape is now in the state $(1-p, p)$, and is uncorrelated from the state in the extractor's memory.
Thus far, no exchange of heat or work has been required.

For the final stage of extraction, however, we must again employ the protocol in lemma~\cref{lemma:TwoLevelLandauer}, and in conjunction with heat bath at temperature $T_E$ quasistatically reset the state on the tape from $(1\!-\!p,p)$ to $X_{\rm dflt}$.
This protocol required a work exchange of $\kB T_E \left(h(p)- \Ent{X_{\rm dflt}}\right)$, corresponding to $-W^1_{\rm out}$ from eq.~\eqref{eq:TapeFreeEnergy}, and concludes the extraction.

In summary, if we now consider a cycle of generation followed by extraction (as above) and set $T_G=T_E$, then the total of all work exchange terms is a net dissipation over the entire cycle is given by $\kB T h(p)$ -- the value of $\mathcal{W}^1_{\rm diss} (S)$ as predicted in eq.~\eqref{eq:MinimumDissipationEq}.
We have thus established constructive protocol in the trajectory formalism for manipulating the pattern at a work cost that saturates the bounds given in the top row of table \cref{table:PerturbedCoinCosts}.

Alternatively, if we choose $p < q < \frac{1}{2}$ [where $p$ is the parameter from the perturbed coin process, and $X_{\rm dflt}=(q,1-q)$] and choose $T_E > T_G$, the above cycle can now function as a heat engine, as drawn in \cref{fig:SimExtCycle}.
The efficiency of this engine is directly calculated
\begin{align}
\eta & = 
 \dfrac{\kB T_E \left[\Ent{X_{\rm dflt}} - h(p)\right] - \kB T_G \Ent{X_{\rm dflt}}}{\kB T_E \left[ \Ent{X_{\rm dflt}} - h(p)\right]} \nonumber \\
 & = 1 - \frac{T_G}{T_E} - \frac{T_G\; h(p)}{T_E \left[\Ent{X_{\rm dflt}} - h(p) \right]},
\end{align}
and exactly matches the bound in eq.~\eqref{eq:CarnotEff} previously derived using information theory.
For case of the perturbed coin, we thus conclude that only the simplest pattern (where $p=0$ such that all states are the same, or $p=1$ such that all states perfectly alternate) will achieve the Carnot efficiency.

\newpage
An emergent proof (one that does not already accept the second law as true) that the protocols detailed in this example are optimal would well beyond the scope of this illustrative example%
\footnote{
If one takes a prescriptive ``bottom-up'' approach (e.g.,~\cite{MandalJ12}) rather than a proscriptive ``top-down'' approach (e.g.,\ this article) to pattern thermodynamics, one can specify a particular process and consider what it does, rather than consider a particular task (here, a pattern manipulation defined only by its inputs and outputs) and work out how best to do it. 
The prescriptive approach typically avoids the question of optimality, as it tends to provide upper bounds (``there is a process at least this efficient'') rather than lower bounds (``no process can be more efficient than this'').}%
, as it requires a tricky optimization over {\em all} possible operations that could be done on the joint tape-memory system (including allowing for an arbitrary amount of ancillary memory, etc.). 
On the other hand, this is where the power of the information-theoretic results derived in our article can be demonstrated: 
 we may assert with confidence that the above protocol {\em is} optimal, since it saturates our bounds.
If there were a protocol using prescient memory that is more work-efficient than this (in the trajectory formalism, or indeed any other framework), then Landauer's principle could not hold, and this would have drastic impact on our understanding of the second law -- at least as to how it applies to the particular physical framework employed.
Thus, if we have faith in the second law and how it has been applied within the physical mechanism, then we can be content to halt our search for a better mechanism here.


\end{document}